\documentclass[reprint,amsmath,amssymb,aps]{revtex4-2}
\usepackage{amsmath}
\usepackage{tipa}
\usepackage{bbm}
\usepackage{txfonts}
\usepackage{graphicx}
\usepackage{dcolumn}
\usepackage{comment}
\usepackage{bm}
\usepackage[mathscr]{eucal}
\usepackage{amssymb}
\usepackage{amsfonts}
\usepackage{latexsym}
\usepackage{color}
\usepackage{graphicx}
\usepackage{floatrow}
\usepackage{xfrac}
\usepackage{mathrsfs} 
\usepackage[colorlinks=true,linkcolor=blue,citecolor=blue,urlcolor=blue,]{hyperref}
\usepackage[doipre={doi:~}]{uri}

\begin{document}

\bibliographystyle{apsrev4-2.bst}

\title{Emergent spatial symmetry and inter-manifold avoided crossing of spin-1 lattice gas in the intermediate interaction regime}
\author{Xue-Ting Fang$^{1}$}
\author{Kun Yuan$^{1}$}
\author{Lushuai Cao$^{1}$}\email[E-mail: ]{lushuai\_cao@hust.edu.cn}
\author{Zhong-Kun Hu$^{1,2}$}\email[E-mail: ]{zkhu@hust.edu.cn}

\affiliation{$^1$National Gravitation Laboratory, MOE Key Laboratory of Fundamental Physical Quantities Measurement, and School of Physics, Huazhong University of Science and Technology, Wuhan 430074, People's Republic of China\\
$^2$Wuhan Institute of Quantum Technology, Wuhan 430206, People’s Republic of China} 

\date{\today}

\begin{abstract}
We investigate the low-filling spin-1 lattice gas in the intermediate interaction regime, in which 
the atom-atom interaction allows the decomposition of the system into the coupled spin and charge sectors,
with lower energetical detuning between the two sectors than in the strong interaction regime.
The low-lying eigenstates are grouped into different manifolds due to the decomposition, 
and are endowed with the emergent spatial inversion symmetry separately in the spin and charge sectors,  
which induces hidden correlations and affects the spin distribution of the system. 
The lowered energetical detuning between the two sectors activates the inter-sector coupling, 
and overlaps different manifolds in the eigenenergy spectrum, 
which leads to the crossings of eigenstates from different manifolds.
The inter-sector coupling between the spin and charges is then witnessed by the the inter-manifold avoided crossings,
which takes place between accidentally degenerate eigenstates of the same symmetry parity.
Our work reveals the enhanced coupling effects between the spin and charge dopants of the spinor lattice
gas in the intermediate interaction regime.
\end{abstract}


\maketitle

\section{Introduction}
The ultracold spinor lattice gas, both bosonic\cite{Davis1995,Ketterle2002,Morsch2006,Becker2008,Bloch2008} and fermionic\cite{DeMarco1999,Zwierlein2005,Giorgini2008,Murmann2015,Omran2015}, 
is composed of ultracold atoms confined in the optical lattice, 
of which the internal states contribute to the spin degree of freedom (DoF). 
The flexible control \cite{Greiner2002,Bloch2005,Bloch2008,Sidorenkov2013} of the lattice potential and interaction,
as well as the rich choice of the spin states have enabled the spinor 
lattice gas as a versatile platform for the quantum simulation\cite{Bloch2005,Bernien2017, Gross2017, Chiu2019, Sun2018,   Wang2021,Zhou2022}, computation\cite{Duan2003,Schneider2012,Shui2021,Graham2022,McDonnell2022,Fang2022,GonzalezCuadra2023} 
and metrology\cite{  Swallows2011,Pezze2018,Fraisse2019,Liu2022,Aeppli2024}.
Particularly, the spinor lattice gases in the strong interaction regime have been extensively explored for the quantum
simulation of various spin lattice models, such as the Heisenberg\cite{Murmann2015,Deuretzbacher2017,Chung2021},
the bi-quadratic spin lattices\cite{GarciaRipoll2004,Chung2009,Zhu2018}.
The magnetic phases of the spinor lattice gas of different spins have been theoretically\cite{Katsura2013,PhysRevLett.102.140402,PhysRevLett.122.053401,PhysRevA.97.023628},
and experimentally\cite{Greiner2002,Zeiher2017,Chiu2019,Sun2021} investigated.
Moreover, the  spinor lattice gas of non-uniform fillings in the strong interaction regime can be mapped to the
coupled spin chain and dopants\cite{Kleine2008,Hond2022},
and the coupling between the spin chain and the charge- and hole-dopant can be engineered, 
which can be applied for the simulation of the quantum magnetism and high-temperature superconductors.

The spinor lattice gas also permits the engineering of the emergent symmetry for the spin DoF, through tuning the
lattice Hamiltonian, e.g., the spin-dependent lattice potential and the interaction strength, 
as well as loading atoms of different species to the lattice.
For instance, tuning the spin-dependent interaction strength of the lattice atoms can enlarge the intrinsic symmetry to 
the emergent ones for the internal DoF, such as for the spin-1 lattice gas from $\mathrm{SU}(2)$ to $\mathrm{SU}(3)$ \cite{Yip2003,Imambekov2003,Rizzi2005,Guan2008},
and the spin-3/2 lattice gas from the exact $\mathrm{SO}(5)$ to $\mathrm{SO}(7)$,
$\mathrm{SU}(4)$ and $\mathrm{SO}(5)\times \mathrm{SU}(2)$ at different critical interaction strengths\cite{Wu2003,WU2006}.
The spin-2 lattice gas can also exhibit the emergent $\mathrm{SO}\left(3\left(5\right)\right)$ and 
$\mathrm{SU}\left(3\left(5\right)\right)$ symmetries as the interaction strength of different scattering channels varies\cite{Chen2015,Yang2019}. 
The $\mathrm{SU}(N)$ symmetry endows novel properties to the spinor lattice gas\cite{Gorshkov2010},
and has been engineered for the nuclear spins of alkaline-earth atoms\cite{Cazalilla2009,Messio2012,Cazalilla2014,Zhang2014,Taie2022},
which can be further enlarged to $\mathrm{SU}(N)\times \mathrm{SU}(2)$ by combining the
nuclear and electronic spins of the lattice gas of alkaline-earth atoms\cite{Gorshkov2010}. 
Given that the current focus of the emergent symmetry is on the spin states,
it is then intriguing to ask whether the emergent symmetry can be extended to the external, i.e., spatial DoF, 
with new effects on the spinor lattice gas.

In this work, we investigate the low-filling spin-1 lattice gas in the intermediate interaction regime,
which lies between the strong and weak interaction regimes.
The intermediate interaction regime inherits from the strong interaction regime that the low-lying eigenstates mainly reside in the single-occupation Hilbert subspace, 
and can be grouped into different manifolds,
which are spanned by the basis states with each lattice site mostly occupied by one atom. 
As the interaction varies from the strong to the intermediate regime, however,
these manifolds evolve from being well gapped to being energetically overlapping in the eigenenergy spectrum, 
which leads to the accidental degeneracy between eigenstates belonging to different manifolds.
The accidental degeneracy is manifested as crossings of the related eigenstates in the spectrum as a 
function of the interaction strength.
Moreover, the low-lying eigenstates possess higher symmetries than the complete Hamiltonian, 
i.e., the emergent symmetry, and consequently the hidden correlations, 
which cannot be directly observed in the complete Hamiltonian of the system.  
The symmetry of the accidentally degenerate eigenstates also determines whether the corresponding crossing
is a direct or an avoided one.
Our work demonstrates that the intermediate interaction regime of the spinor lattice gas holds the promise to engineer both the
emergent symmetries and the coupling effects between the spin lattice and the charge dopants, which could reveal new properties
in the doped spin lattice systems. 

This paper is organized as follows: In Sec.\ref{II} we introduce the Hamiltonian of 1D spin-1 Bose gas and present the derivation of low-energy effective Hamiltonian. In Sec.\ref{III} we present the numerically calculated energy spectrum and the symmetry-driven energy-level crossings between different manifolds. Finally, a brief discussion and conclusion are given in Sec.\ref{IV}.

\section{Setup and decomposition of the spin-1 lattice gas}\label{II}
We consider the spinor lattice gas composed of $N$ spin-1 bosonic atoms confined in the 
one-dimensional optical lattice of $L$ sites, with $N<L$, i.e., a filling lower than unity.
Under the tight-binding approximation, the Hamiltonian of the spin-1 lattice gas is given by:
\begin{equation}
\begin{aligned}
\hat H _{HB}&=  - t\sum_{ < i,j > ,\alpha } {\hat a_{i,\alpha }^\dag {{\hat a}_{j,\alpha }}}  + {c_0}\sum_{i,\alpha ,\beta } {\hat a_{i,\alpha }^\dag \hat a_{i,\beta }^\dag {{\hat a}_{i,\beta }}{{\hat a}_{i,\alpha }}}  \\
&+{c_2}\sum_{i,\alpha \beta \gamma \upsilon } {\hat a_{i,\alpha }^\dag {\vec{\sigma} _{a\beta }}{{\hat a}_{i,\beta }}\hat a_{i,\gamma }^\dag {\vec{\sigma} _{a\beta }}{{\hat a}_{i,\upsilon }}} ,
\end{aligned}
\end{equation}
in which $\hat a_{i,\alpha }^{\left(\dag\right)}$ is the annihilation (creation) operator of one atom of the spin state $\alpha$ in the
$i$-th lattice site, with $\alpha\in\{1,0,-1\}$.
$\vec \sigma  = \left( {{\sigma _x},{\sigma _y},{\sigma _z}} \right)$ is the spin-1 matrix representation.
Besides the hopping of the atoms in the lattice given in the first term, the second and third terms of $\hat H_{\rm{HB}}$ denote
the symmetric and asymmetric atom-atom interactions, respectively.
The pair interaction coefficients are 
${c_0} = {{4\pi {\hbar ^2}\left( {{a_0} + 2{a_2}} \right)} \mathord{\left/ {\vphantom {{4\pi {\hbar ^2}\left( {{a_0} + 2{a_2}} \right)} {3M}}} \right. \kern-\nulldelimiterspace} {3M}}$ and 
${c_2} = {{4\pi {\hbar ^2}\left( {{a_2} - {a_0}} \right)} \mathord{\left/ {\vphantom {{4\pi {\hbar ^2}\left( {{a_2} - {a_0}} \right)} {3M}}} \right. \kern-\nulldelimiterspace} {3M}}$, 
with ${a_0}\left( {{a_2}} \right)$ denoting the s-wave scattering length for two spin-1 atoms 
in the combined symmetric channel of total spin 0 (2).
 
In the intermediate and strong interaction regimes, the low-lying eigenstates primarily occupy the Hilbert subspace
spanned by the single-occupation states, of which each site is mostly occupied by one atom.
The single-occupation states can then be written as
$|\sigma_{i_1},\sigma_{i_2},\cdots,\sigma_{i_N}\rangle$,  
of which ${i_1} < {i_2} <  \cdots  < {i_N}\in [1,L]$ denotes the index of the sites occupied by the atoms, 
and $\sigma_{i_n}\in \{-1,0,1 \}$ refers to the spin state of the atom in the $i_n$-th site.
Under this truncation, the spinor lattice gas can be mapped to a spin chain coupled to hard-core hole dopants, 
which can be viewed as the decomposition of the single-occupation Hilbert subspace into a spin sector and a 
charge sector\cite{Hilker2017,Vijayan2020}. 
In the decomposition, the spin sector is composed of a chain of $N$ spins in the squeezed space with the elimination of holes, 
and the charge sector is formed by $\left(L-N\right)$ hard-core holes in an $L$-site lattice,
for which the corresponding basis-state transformation can be given as:
\begin{equation}\label{basis_trans}
\begin{aligned}
|\sigma_{i_1},\sigma_{i_2},\cdots,\sigma_{i_N}\rangle =
         {\left| {{\sigma _1},{\sigma _2}, \cdots ,{\sigma _N}} \right\rangle_s} \otimes {\left| h_1<h_2<\cdots<h_{L-N} \right\rangle_h}.
\end{aligned}
\end{equation}
In the above equation,  ${\left| {{\sigma _1},{\sigma _2}, \cdots, {\sigma _N}} \right\rangle_s}$
and $\left| h_1<h_2<...<h_{L-N} \right\rangle_h$ denote the basis states in the spin and charge sectors, respectively, 
of which $h_i$ denotes the location of the $i$-th hole-occupied site.

\begin{figure}[t]
\includegraphics[trim=0 5 0 5,width=1\textwidth]{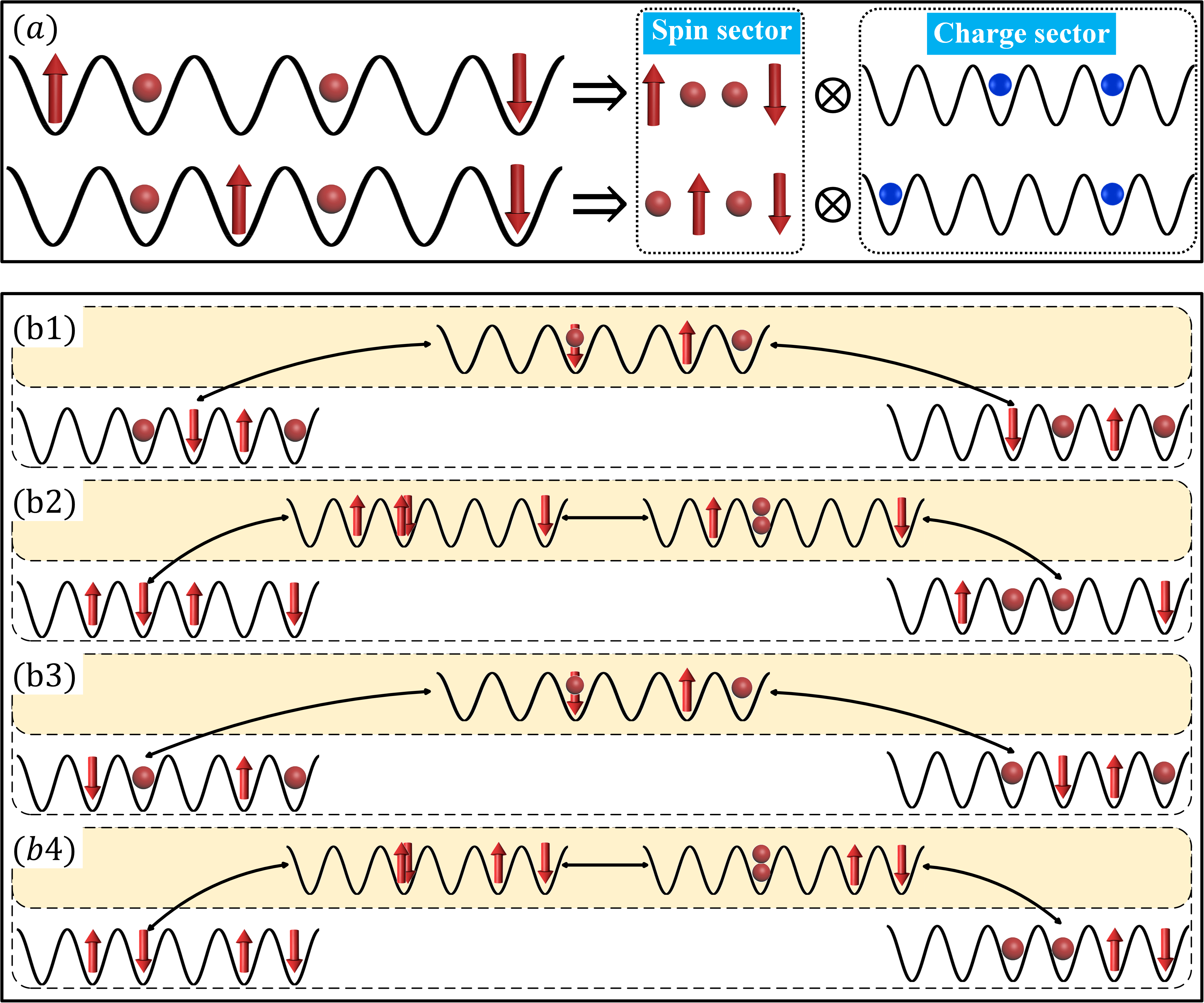}
\caption{\label{fig-configuration} (a) Sketch of the decomposition of the spinor lattice gas within the single-occupation Hilbert subspace
to the spin and charge sectors. (b) The higher-order perturbative channels of the spin-spin exchange (b1) and scattering (b2) interactions,
and spin-spin exchange (b3) and scattering  (b4) interactions associated with the next-nearest-neighbor hopping of the hole.}
\end{figure}

Following the decomposition, the spinor lattice gas can be described by the 
$\rm{t-J}$ Hamiltonian\cite{Emery1990,Ogata1991,Grusdt2020,Fang2022,Hirthe2023}, as 
${\hat H_{{\rm{t-J}}}} = {\hat H_{{\rm{charge}}}} + {\hat H_{{\rm{spin}}}} + {\hat H_{{\rm{sp-ch}}}}$:
\begin{equation}
\begin{split}
{{\hat H}_{{\rm{charge}}}} &=  - t\sum_{i = 1}^L {\hat c_i^\dag {{\hat c}_{i + 1}} + {\rm{H}}.{\rm{c}}.} ,\\
{{\hat H}_{{\rm{spin}}}} 
                          &=  J\sum_{j = 1}^{N - 1} \hat H_{BB}(j),\\
{{\hat H}_{{\rm{sp-ch}}}} &=  - J\sum_{j = 1}^{N - 1} \hat H_{BB}(j) \left(1-\Delta\left(j\right)\right) \\
                          &+ J\sum_{j = 1}^{N - 1}\hat H_{BB}(j)\Delta\left(j\right)
                                            \left(\hat c_{\eta(j)}^\dag \hat c_{\eta(j)+2}+\rm{H}.\rm{c}. \right).                          
\end{split}
\end{equation}
In ${\hat H_{{\rm{t-J}}}}$, $\hat c_i^{\left( \dag  \right)}$ refers to the annihilation (creation) operator of holes at
the $i$-th site of the lattice in the charge sector, 
and $\hat H_{BB}(j)=\left( \alpha \left( \hat{ \vec{\sigma}}_j \cdot \hat{\vec{\sigma}}_{j+1} \right)^2
+\hat{\vec{\sigma}}_j \cdot \hat{\vec{\sigma}}_{j+1}+\beta \right)$ 
describes the  bilinear-biquadratic  spin-spin interaction between the $j$- and $\left(j+1\right)$-th spins in the spin sector,
with $\hat{\vec{\sigma}}_j=\left(\hat \sigma_{j,x},\hat \sigma_{j,y},\hat \sigma_{j,z}\right)$ 
denoting the spin-1 operators of the $j$-th spin in the chain of the spin sector.
The coefficients and constants in ${\hat H_{{\rm{t-J}}}}$ are derived through the second-order perturbation to $\hat H_{\rm{HB}}$
as $J = {t^2}/ {\left( {{c_0} + {c_2}} \right)}$, as well as $\alpha={c_0}/{(2c_2-c_0)}$ and $\beta={2c_2}/{(2c_2-c_0)}$. 
In the above equations, ${\hat H}_{{\rm{charge}}}$ describes the hopping of the hole in the lattice of the charge sector.
The spin sector turns out to a bilinear-biquadratic spin (BBS) chain described by ${\hat H}_{{\rm{spin}}}$.

${\hat H}_{{\rm{sp-ch}}}$ presents the coupling between the spin and charge sectors, in which
$\Delta\left(j\right)=1\left(0\right)$ refers to that 
the atoms corresponding to the $j$- and $\left(j+1\right)$-th spins of the BBS chain are (not) remaining neighbors
in the original spinor lattice system.
$\Delta\left(j\right)$ can then be derived as $\Delta\left(j\right)=\delta \left( \eta(j)+1, \eta(j+1) \right)$,
where $\eta(j)$ maps the location of the spin in the BBS chain to that of the corresponding atom
in the original optical lattice, given by 
$\eta(j)= \sum_{\alpha  = 1}^L {\alpha  \cdot \delta \left( {\sum_{\beta  = 1}^{\alpha  - 1} 
{{{\hat n}_\beta }} ,\alpha  - j} \right) \cdot \delta \left( {{\hat n_\alpha },0} \right)}$,
with $\hat n_\alpha=\hat c^\dag_\alpha\hat c_\alpha$ the density operator of the holes.
The first term in ${\hat H}_{{\rm{sp-ch}}}$ describes that the spin-spin interaction between the
$j$- and $\left(j+1\right)$-th spins is prevented by the appearance of hole(s) in between the corresponding atoms of
the two spins in the original spinor lattice, and
the second term refers to the fact that the spin-spin interaction can be accompanied with the next-nearest neighbor
hopping of the hole in the charge sector. 

Figure \ref{fig-configuration} illustrates decomposition of the spinor lattice gas into the spin and charge sectors, and the 
dominant second-order perturbation channels for $\hat H_{\rm{t-J}}$. The one-to-one correspondence between the single-occupation
states of the original spinor lattice gas and the basis states of the spin and charge sectors is exemplified in 
Fig. \ref{fig-configuration} (a).
The second-order processes of the spin-spin exchange and scattering interactions given in $\hat H_{\rm{spin}}$
are illustrated in Fig. \ref{fig-configuration} (b1) and (b2), respectively.
Figures \ref{fig-configuration} (b3) and (b4) sketch the spin-spin exchange and scattering interactions associated with the next-nearest 
neighbor hopping of the neighboring hole given in the second term of $\hat H_{\rm{sp-ch}}$, respectively. 

$\hat H_{\rm{spin}}$ and $\hat H_{\rm{charge}}$ commute with different symmetry operators from $\hat H_{\rm{sp-ch}}$,
which underlies the emergent spatial symmetry of the low-lying eigenstates, as will be discussed in the next section.
The BBS chain in the spin sector exhibits the spin-flipping and spatial inversion symmetries, 
and correspondingly $\hat H_{\rm{spin}}$ commutes with the symmetry operators of 
${\hat {\cal S}}_{sp}$ and ${{\hat {\cal R}}_{sp}}$, of which ${\hat {\cal S}}_{sp}$ flips each spin in the
BBS chain between states $|1\rangle$ and $|-1\rangle$, and ${{\hat {\cal R}}_{sp}}$ exchanges the spin states of
the $n$-th and $\left(N-n+1\right)$-th spins, with $n\in[1,N]$.
The charge sector also exhibits spatial inversion symmetry of the holes, with the symmetry operator 
${{\hat {\cal R}}_{ch}}$, which exchanges the hole-occupation states between the $l$-th and $\left(L-l+1\right)$-th sites 
in the hole sector, with $l\in[1,L]$.
The coupling term $\hat H_{\rm{sp-ch}}$ and consequently the complete $\hat H_{\rm{t-J}}$ commute with ${\hat {\cal S}}_{sp}$ and 
${\hat {\cal R}}_{sp}\times{{\hat {\cal R}}_{ch}}$, but not with the individual ${\hat {\cal R}}_{sp}$ or ${{\hat {\cal R}}_{ch}}$.
Given that $\hat H_{\rm{sp-ch}}$ plays the role of a perturbation in the intermediate interaction regime,
the low-lying eigenstates, which are not in accidental degeneracy, present well-defined parity with respect to
${{\hat {\cal R}}_{sp}}$ and ${{\hat {\cal R}}_{ch}}$, and exhibit a higher symmetry than $\hat H_{t-J}$, 
i.e., the appearance of the emergent spatial inversion symmetry.

\begin{figure}[t]
\includegraphics[trim=0 5 5 0,width=1\textwidth]{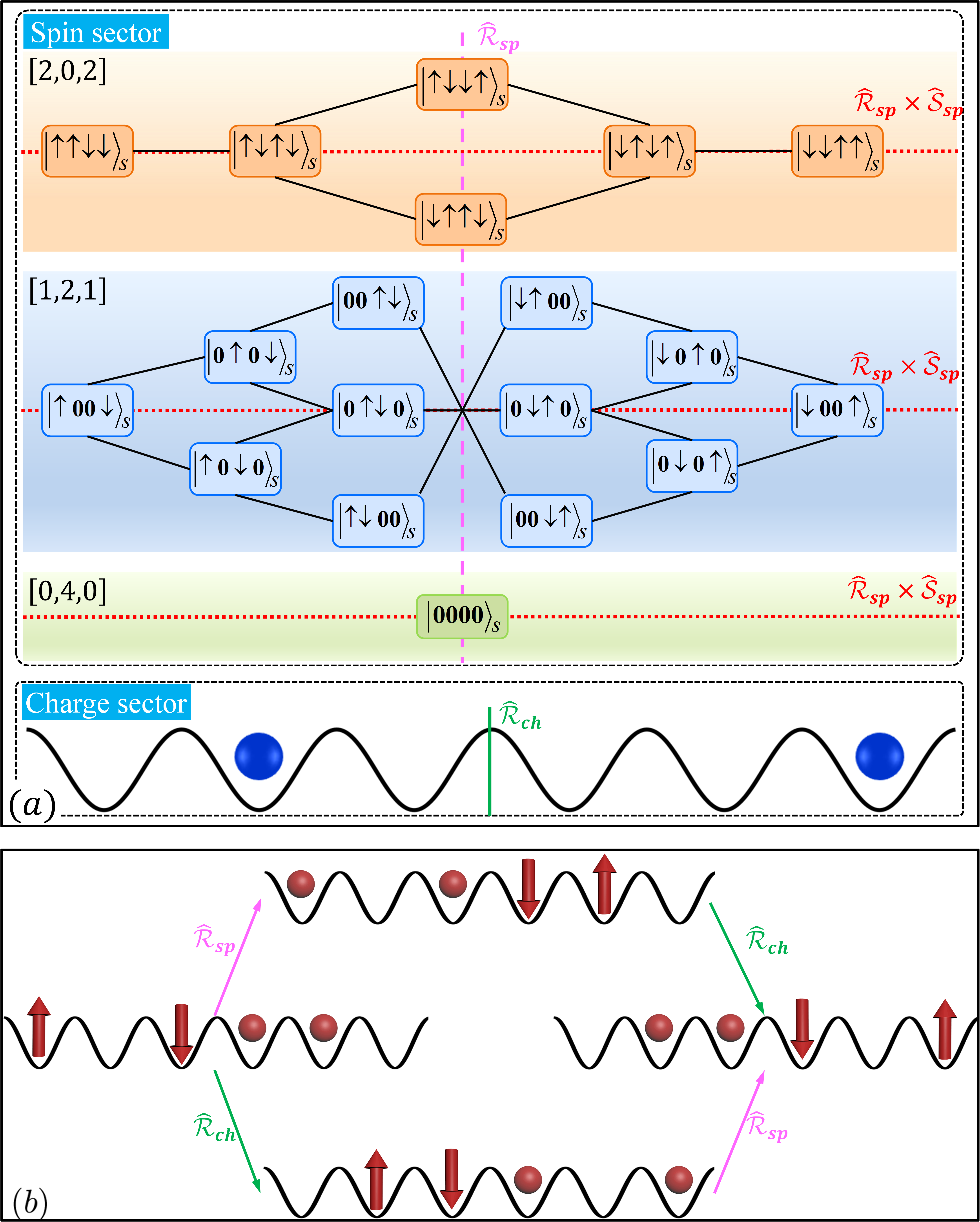}
\caption{\label{fig-symmetry}(a) Sketch of the spin flipping and the spatial inversion symmetry of the BBS chain as the rotations
in the configuration space of the spin sector (upper panel), and the spatial inversion symmetry of the charge sector (lower panel).
The magenta dashed and red doted lines in the spin-sector configuration space denotes the rotation axis of the symmetric operation
$\hat {\cal R}_{sp}$ and $\hat {\cal S}_{sp}\times\hat {\cal R}_{sp}$, respectively in the upper panel, and green solid line in the
lower panel denotes the rotation axis of $\hat {\cal R}_{ch}$.
(b) The effect of the spatial inversion operation of the spin and charge sector, as well as their combination, 
in the original spinor lattice system.
}
\end{figure}

Figure \ref{fig-symmetry} illustrates the symmetries in the spin and charge sectors, as well as the original spinor lattice gas,
for a spinor lattice gas with $\left(N,L\right)=\left(4,6\right)$.
In the upper panel of Fig. \ref{fig-symmetry} (a), the symmetry operators ${{\hat {\cal S}}_{sp}}$ and ${{\hat {\cal R}}_{sp}}$
are sketched in the configuration space of the subsectors $[2,0,2]$, $[1,2,1]$ and $[0,4,0]$, 
where the subsector $[n_1,n_0,n_{-1}]$ is spanned by the basis states of the BBS chain with the number of spins in state $\sigma$ as $n_\sigma$. 
In the configuration space, the basis states within each subsector are coupled by the spin-spin exchange interaction, 
illustrated by the solid lines in the figure, 
and the inter-subsector coupling is induced by the spin-spin scattering interaction, which is not shown in the figure.
The configuration space of each subsector presents the rotation symmetry with respect
to the vertical and horizontal symmetry axes, denoted by the magenta dashed and red dotted lines in the figure, respectively.
It turns out that the rotation symmetry around the vertical and horizontal axes is the 
symmetry operations of $\hat {\cal R}_{sp}$ and $\hat {\cal S}_{sp}\times\hat {\cal R}_{sp}$,
respectively. The basis states along the vertical and horizontal rotation axes are then manifested as the high-symmetry points of the
corresponding symmetry operations.
The space inversion symmetry in the charge sector is shown in the lower panel of Fig. \ref{fig-symmetry} (a), where the green solid
line indicates the rotation axis of the symmetry operation.

In the original spinor lattice system, as sketched in Fig. \ref{fig-symmetry} (b), $\hat {\cal R}_{sp}$ induces the spatial inversion
in the spin distribution among the atoms, leaving the location of the atoms (holes) not affected, 
and oppositely, $\hat {\cal R}_{ch}$ only inverses the location of the atoms (holes), with
the distribution of the spin states among the atoms not changed. The intrinsic spatial inversion symmetry of the spinor lattice
system corresponds to the combination of the two operations, with the symmetry
operator as ${\hat {\cal R}}_{sp}\times{{\hat {\cal R}}_{ch}}$. 
Similar to the explicit correlation between a single-occupation basis state $|\alpha\rangle$ and its symmetric counterpart  
${\hat {\cal R}}_{sp}\times{{\hat {\cal R}}_{ch}}|\alpha\rangle$, the emergent spatial inversion symmetry in the two sectors
can induce the additional correlation of $|\alpha\rangle$ with ${\hat {\cal R}}_{sp}|\alpha\rangle$ and ${{\hat {\cal R}}_{ch}}|\alpha\rangle$,
which is hidden from the complete Hamiltonian of $\hat H _{HB}$ and ${\hat H_{{\rm{t-J}}}}$.


\section{Numerical results}\label{III}
\begin{figure}[t]
\includegraphics[trim=25 5 50 5,width=1\textwidth]{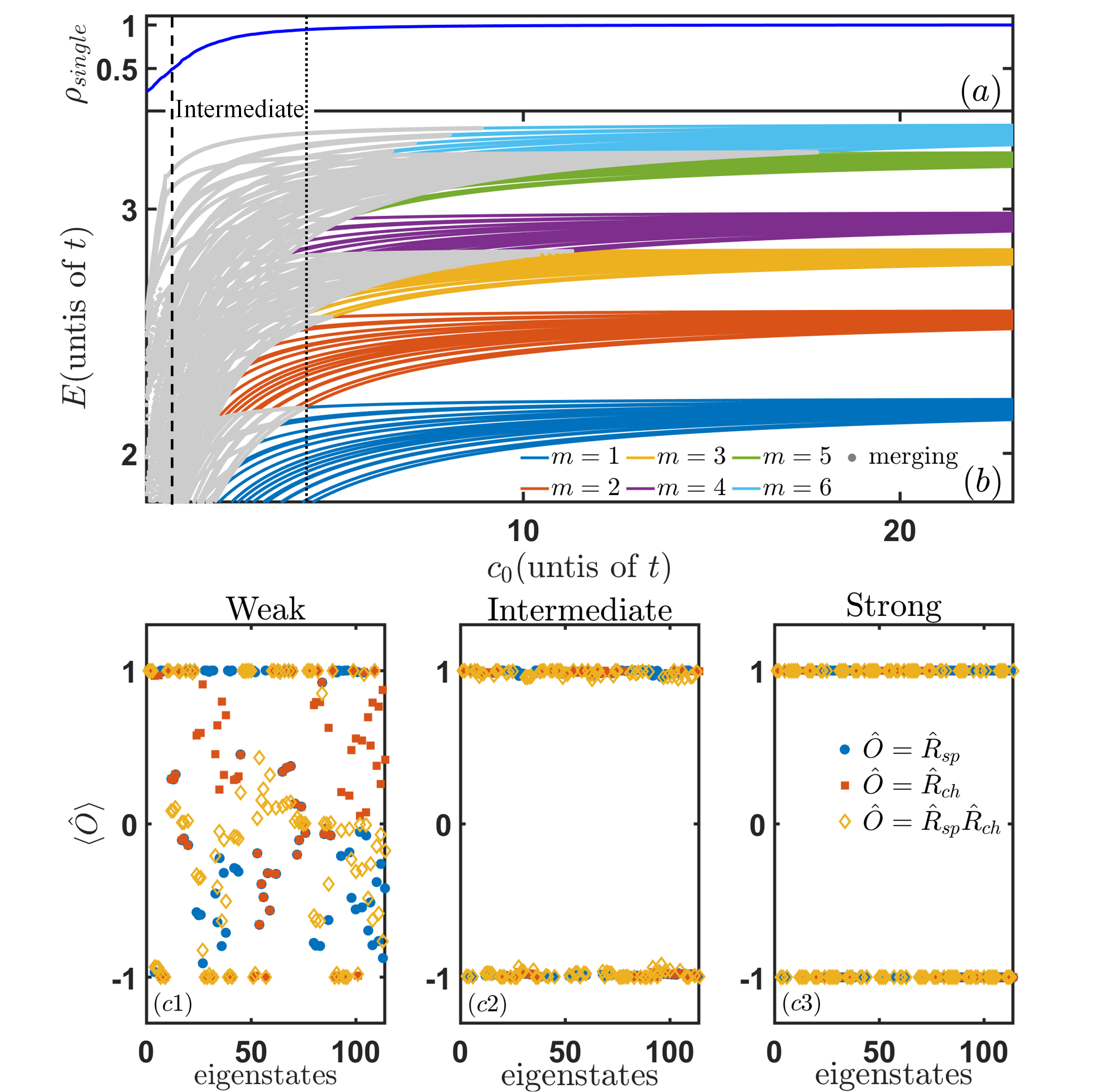}
\caption{\label{fig-energy} (a) The total probability of the single-occupation basis averaged over the low-lying eigenstates
and (b) the eigenenergy spectrum as a function of $c_0$ for the spinor lattice gas with $\left( N=4,L=6,S_z=0\right)$. 
(c) The normalized expectation of the symmetry operators $\langle{{\hat {\cal R}}_{sp}\rangle}$ (dots), 
$\langle{{\hat {\cal R}}_{ch}}\rangle$ (squares), 
and $\langle{{\hat {\cal R}}_{sp}\times{{\hat {\cal R}}_{ch}}\rangle}$ (diamonds) for the low-lying eigenstates in
the weak (c1), intermediate (c2) and strong interaction regimes(c3).}
\end{figure}

We numerically investigate the low-lying eigenstates of the finite spin-1 lattice gas with $\left( N = 4,L = 6 \right)$ in
this section, which are done by the exact diagonalization of $\hat H_{\rm{HB}}$ in the complete Hilbert space, 
with no truncation to the single-occupation subspace. The numerical simulation scans a wide range of the interaction
strength $c_0$ from the weak to approaching the Tonks-Girardeau regime, with $c_2=0.1c_0$ fixed in all the calculations. 
Figure \ref{fig-energy}(a) provides the total probability of the single-occupation states $\rho_{single}\left(c_0\right)$
averaged over the low-lying eigenstates, with $\rho_{single}\left(c_0\right)=\sum_{\Psi}\langle \Psi|\hat P_s|\Psi\rangle/N_{tot}$, 
where $|\Psi\rangle$ sums over all the low-lying eigenstates and $\hat P_s$ is the projector to the singly-occupied Hilbert subspace, 
with $N_{tot}$ denoting the total number of low-lying eigenstates. 
Figure \ref{fig-energy}(b) shows the eigenenergy spectrum of the low-lying eigenstates versus $c_0$.

In the strong interaction regime, which has been extensively investigated, Figs. \ref{fig-energy} (a) and (b) reproduce the well
known results that the low-lying eigenstates are well residing in the single-occupation Hilbert subspace with 
$\rho_{single}\left(c_0\right)$ approaching unity, and are grouped into different manifolds,
which are well-gapped in the eigenenergy spectrum.
In Fig. \ref{fig-energy}, different manifolds are marked with different colors in the strong interaction regime.
Given that the single-occupation Hilbert subspace can be decomposed into the spin and charge sectors,
the multi-manifold formation of the low-lying eigenstates can be attributed to the energetical detuning between the two sectors,
which is due to $t \gg J$ in the strong interaction regime. The energetical detuning suppresses the coupling between the two sectors,
and the low-lying eigenstates can then be well approximated by the direct product of the eigenstates of 
the two sectors, i.e., $|m\rangle_{ch}\times|k\rangle_{sp}$, where $m$ and $k$ are the eigenstate index of
$\hat H_{charge}$ and $\hat H_{spin}$, respectively, and are manifested as the good quantum numbers of the low-lying eigenstates.
The low-lying eigenstates belonging to the same manifold share the same quantum number $m$, and the energy gaps between
different manifolds are determined by the energy scale of the charge sector $t$. In the finite system, 
there also exist the mini-gaps between eigenstates in the same manifold due to the finite-size effect,
and the width of mini-gaps, as well as that of the manifolds, are dependent on the energy scale of the spin sector $J$.
In the weak interaction regime, on the other hand, $\rho_{single}\left(c_0\right)$ strongly decreases from unity, indicating
that the low-lying eigenstates are not confined in the single-occupation Hilbert subspace and cannot be grouped into different
manifolds with the good quantum number $m$, as witnessed by the vanishing of the well-gapped multi-manifold structure in the
eigenenergy spectrum.

In Figs. \ref{fig-energy} (a) and (b), there lies the intermediate interaction regime,
where $\rho_{single}\left(c_0\right)$ remains close to unity as in the strong interaction regime, while 
the different manifolds in the eigenenergy spectrum become overlapped.
In the intermediate interaction regime, $\rho_{single}\left(c_0\right)\approx 1$ illustrates that 
the low-lying eigenstates are still concentrating in the single-occupation Hilbert subspace, 
and the decomposition into the spin and charge sectors remains valid. 
The vanishing of the inter-manifold gaps in the eigenenergy spectrum further indicates that the energetical detuning 
between the spin and charge sectors decreases, and would activate 
the coupling between the two sectors in the intermediate interaction regime, which is manifested as the major difference from the strong
interaction regime. In Fig. \ref{fig-energy}(b), in order to visualize the merging between different manifolds,
the color of the eigenenergy plot is changed to gray, once it crosses with another eigenstate from a different manifold,
as $c_0$ decreases. Since there is no transition between the intermediate interaction regime and the weak as well as the strong regime, 
the interval of the intermediate interaction regime is roughly sketched by the blue dashed and red dotted lines
in Figs. \ref{fig-energy} (a) and (b), of which the location is chosen by $\rho_{single}\left(c_0\right)=0.5$ and 
the vanishing of all inter-manifold gaps in the spectrum, respectively.

The decomposition of the spinor lattice gas into the spin and charge sectors, 
with $\hat H_{\rm{spin}}$ and $\hat H_{\rm{charge}}$ commuting with different symmetry operators
from $\hat H_{\rm{sp-ch}}$, can lead to the emergent spatial inversion symmetry of the low-lying eigenstates.
Figures \ref{fig-energy} (c1-c3) examine these emergent spatial inversion symmetry 
for different interaction regimes, by the normalized expectation values of the corresponding symmetry operators,
$\langle{{\hat {\cal R}}_{sp}\rangle}$, $\langle{{\hat {\cal R}}_{ch}}\rangle$, 
which are defined as
$\langle\hat O\rangle=\sum_{\Psi}\langle\Psi|\hat P_s \hat O \hat P_s|\Psi\rangle/\langle\Psi|\hat P_s|\Psi\rangle/N_{tot}$,
with $\hat O\in\{{\hat {\cal R}}_{sp},\hat {\cal R}_{ch}\}$ and $|\Psi\rangle$ taken from the low-lying eigenstates.
$\langle{{\hat {\cal R}}_{sp}\rangle}$ and $\langle{{\hat {\cal R}}_{ch}}\rangle$ verifies the hidden correlation induced by
the emergent symmetry in the spin and charge sectors, respectively, and also measure the parity of the corresponding symmetry operators.
In Figs. \ref{fig-energy}(c1-c3), the normalized expectation $\langle{{\hat {\cal R}}_{sp}\times{{\hat {\cal R}}_{ch}}\rangle}$,
corresponding to the exact spatial inversion symmetry of the complete spinor lattice gas, is also shown for comparison, 
which possesses well defined parity $\pm 1$ in all the three interaction regimes, as shown in the figures.
$\langle{{\hat {\cal R}}_{sp}\rangle}$ and $\langle{{\hat {\cal R}}_{ch}}\rangle$ deviates from unity
in the weak interaction regime, as shown in Fig. \ref{fig-energy}(c1), which indicates that 
the separated spatial inversion operation in the spin and charge sectors leads to no exact symmetry of the complete system.
However, in the intermediate and strong interaction regimes,
$\langle{{\hat {\cal R}}_{sp}\rangle}$ and $\langle{{\hat {\cal R}}_{ch}}\rangle$ restores to well-defined parities $\pm 1$, 
which witnesses the arising of the emergent spatial inversion symmetry in the two sectors 
for the low-lying eigenstates in the two interaction regimes. 
It is also worth mentioning that the emergent spatial inversion symmetries in the spin and charge sectors are perturbative ones, 
since the coupling $\hat H_{\rm{sp-ch}}$, which breaks the emergent symmetries, can still induce higher order corrections
to the low-lying eigenstates.

\begin{figure}[t]
\includegraphics[trim=75 5 50 5,width=1\textwidth]{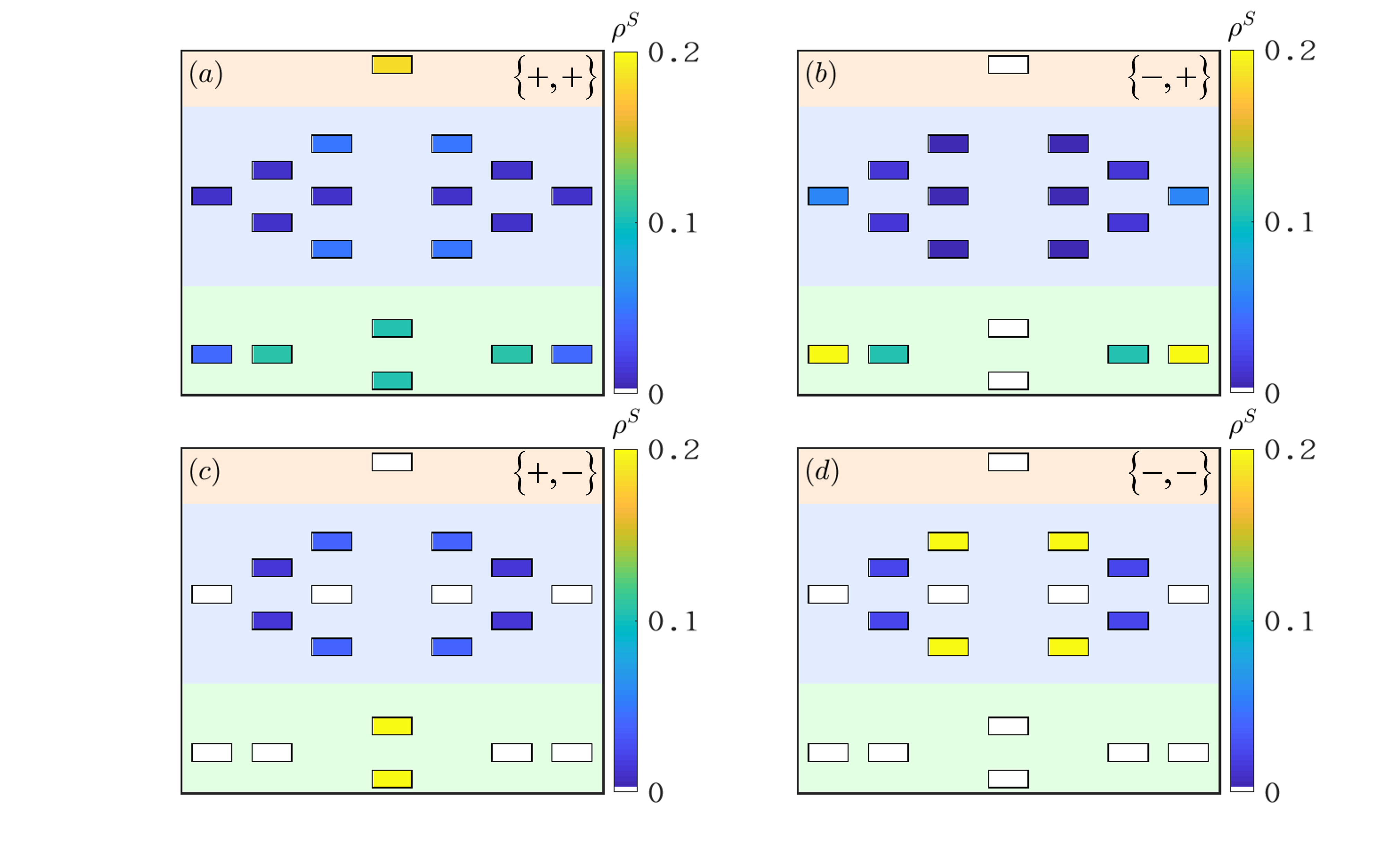}
\caption{\label{fig-rho} Density distributions in the configuration space of the spin sector for the eigenstates with the parities of 
$\{  \langle{{\hat {\cal R}}_{sp}}\rangle, \langle{{\hat {\cal R}}_{sp}\times{{\hat {\cal S}}_{sp}}\rangle} \}$ equaling to 
(a) $\left \{ +,+ \right \} $, (b) $\left \{ -,+ \right \} $, (c)$\left \{ +,- \right \} $ and (d)$\left \{ -,- \right \} $.}
\end{figure}

The emergent spatial symmetry, similar to its counterpart of the exact spatial inversion symmetry,
can affect the density distributions in the corresponding sectors, and could be explored for, e.g.,
the spin state manipulation. In Fig. \ref{fig-rho}, we illustrate the effect of the exact spin flipping symmetry and
the emergent spatial inversion symmetry on the spin density 
distribution, defined as
\begin{equation}
\begin{split}
\rho _{\Psi}^S = \left\langle \Psi \right|{\left. {\vec \sigma } \right\rangle _s}\left\langle {\vec \sigma } \right.\left| \Psi \right\rangle,
\end{split}
\end{equation}
of which $|\Psi\rangle$ is chosen from the low-lying eigenstates and $|\vec \sigma\rangle$ denotes the basis states of the BBS chain in the spin sector.
We focus on the subsectors $[2,0,2]$, $[1,2,1]$ and $[0,4,0]$ as in Fig. \ref{fig-symmetry}, where the symmetry operations
${\hat {\cal R}}_{sp}$ and ${\hat {\cal S}}_{sp}\times{\hat {\cal R}}_{sp}$ correspond to the rotation of the configuration space of each subsector
around the vertical and horizontal symmetric axes, respectively. Figures \ref{fig-rho}(a-d) plot $\rho _{\Psi}^S$ for eigenstates with
different combinations of the symmetry parities 
$\left(\langle{\hat {\cal R}}_{sp}\rangle, \langle{\hat {\cal S}}_{sp}\times{\hat {\cal R}}_{sp}\rangle\right)$. 
It can be found that the spin density distribution is exactly symmetric with respect to the 
vertical and horizontal symmetry axes for all the four parity combinations. Moreover, eigenstates with the negative parities of 
${\hat {\cal R}}_{sp}$ and ${\hat {\cal S}}_{sp}\times{\hat {\cal R}}_{sp}$ avoid occupying the high symmetric points
along the vertical and horizontal axes in the configuration space, respectively. 
In  Fig. \ref{fig-rho}(d), particularly, the spin distribution with
$\left(\langle{\hat {\cal R}}_{sp}\rangle, \langle{\hat {\cal S}}_{sp}\times{\hat {\cal R}}_{sp}\rangle\right)=\left(-1,-1\right)$, 
is restricted to the subsector of $[1,2,1]$ as the negative parities forbid the occupation in subsectors $[2,0,2]$ and $[0,4,0]$, 
which highlights that the (emergent) symmetries can affect and manipulate the spin distribution.

\begin{figure}[t]
\includegraphics[trim=25 5 50 5,width=1\textwidth]{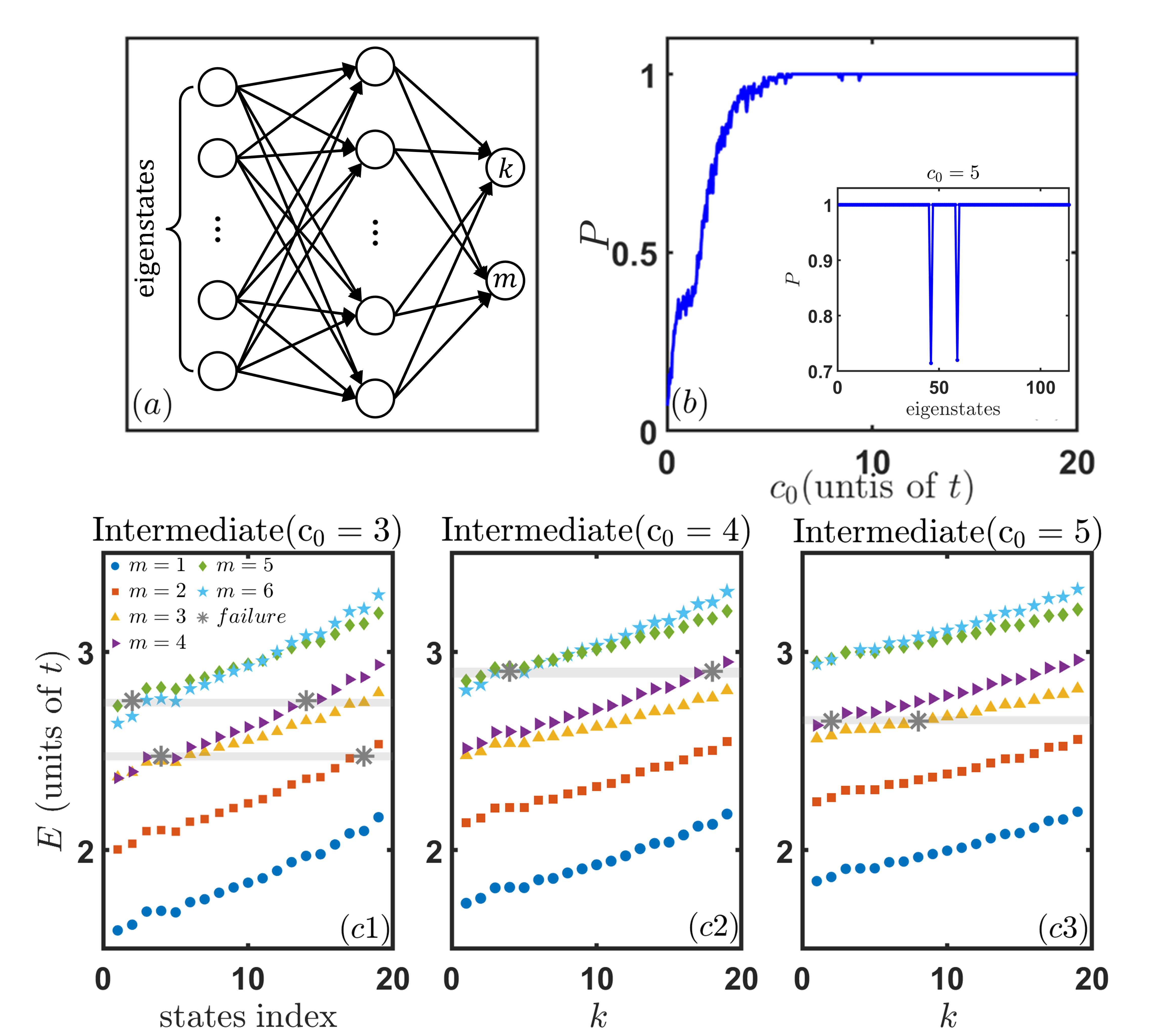}
\caption{\label{fig-ml} (a) The single hidden-layer neural network for the quantum number recognition. 
(b) The averaged success rate of the recognition over the low-lying eigenstates as a function of $c_0$, with the inset showing the success rate of different low-lyng eigenstates for $c_0=5$ in the intermediate interaction regime.
(c) Eigenenergy spectrum as a function of the quantum number $k$ for $c_0=3$ (c1), $c_0=4$ (c2) and $c_0=5$ (c3).
The quantum number $m$ is encoded with the color, and the gray stars refers to the eigenstates of low success rate.}
\end{figure}

In the intermediate interaction regime, the vanishing of the inter-manifold gaps in the eigenenergy spectrum of Fig. \ref{fig-energy}(b)
indicates the weakening of the
energetical detuning between the spin and charge sectors, which can activate the inter-sector coupling. It then 
arises the question of how the coupling affects the low-lying eigenstates in the intermediate interaction regime.
In order to address this question, we firstly examine whether the low-lying eigenstates can still be grouped into different manifolds
even in the absence of the protection by the inter-manifold gaps, 
through the quantum number identification based on the supervised machine learning.
The quantum number identification employs the single-hidden-layer neural network,
as sketched in Fig. \ref{fig-ml}(a), which takes the wavefunction of the eigenstates as the input and 
the identified quantum numbers as the output. The training set is chosen from eigenstates in the strong interaction regime with
labeled quantum numbers $\left(m,k\right)$, and the test set contains the eigenstates in the weak, 
intermediate and also the strong interaction regimes. 
The identification success rate, shown in Fig. \ref{fig-ml}(b), approaches unity in the strong interaction regime,
and presents a sharp drop in the weak interaction regime.
In the intermediate regime, the total success rate remains relatively high, while becomes blurred, 
which indicates that, for one thing, most of the low-lying eigenstates can be recognized with a high 
success rate to the quantum numbers $\left(m,k\right)$,
and for another, there also exist exceptions with the low success rate of recognition. 
In the inset of Fig. \ref{fig-ml}(b), the success rate for the low-lying eigenstates is plotted for a particular $c_0$
in the intermediate interaction regime, which confirms that the recognition success rate remains high for almost all eigenstates
but drops at a pair of exception ones.

In Figs. \ref{fig-ml}(c1-c3), the eigenenergy spectra of different interaction strengths in the intermediate regime
are plotted as a function of the recognized quantum number $k$. In the figures, different colors are used to mark the quantum number
$m$ of the eigenstates, i.e., the manifold index, 
and the eigenstates with low success rate are particularly indicated with gray stars. 
It can be found that the manifolds energetically overlap with each other, 
and eigenstates from different manifolds can become accidentally degenerate, 
even in the presence of the mini-gaps due to the finite-size effect. 
Moreover, the eigenstates plotted in gray stars arise in pairs, and each pair of these eigenstates is degenerate and belongs to 
different manifolds. It then indicates that the low recognition success rate of these eigenstates can be attributed to the fact that
the accidental degeneracy of these eigenstates activates the $\hat H_{sp-ch}$-induced coupling between the spin
and charge sectors, which mixes the eigenstates from different manifolds and leads to the low recognition success rate.

\begin{figure}[t]
\includegraphics[trim=25 5 50 5,width=1\textwidth]{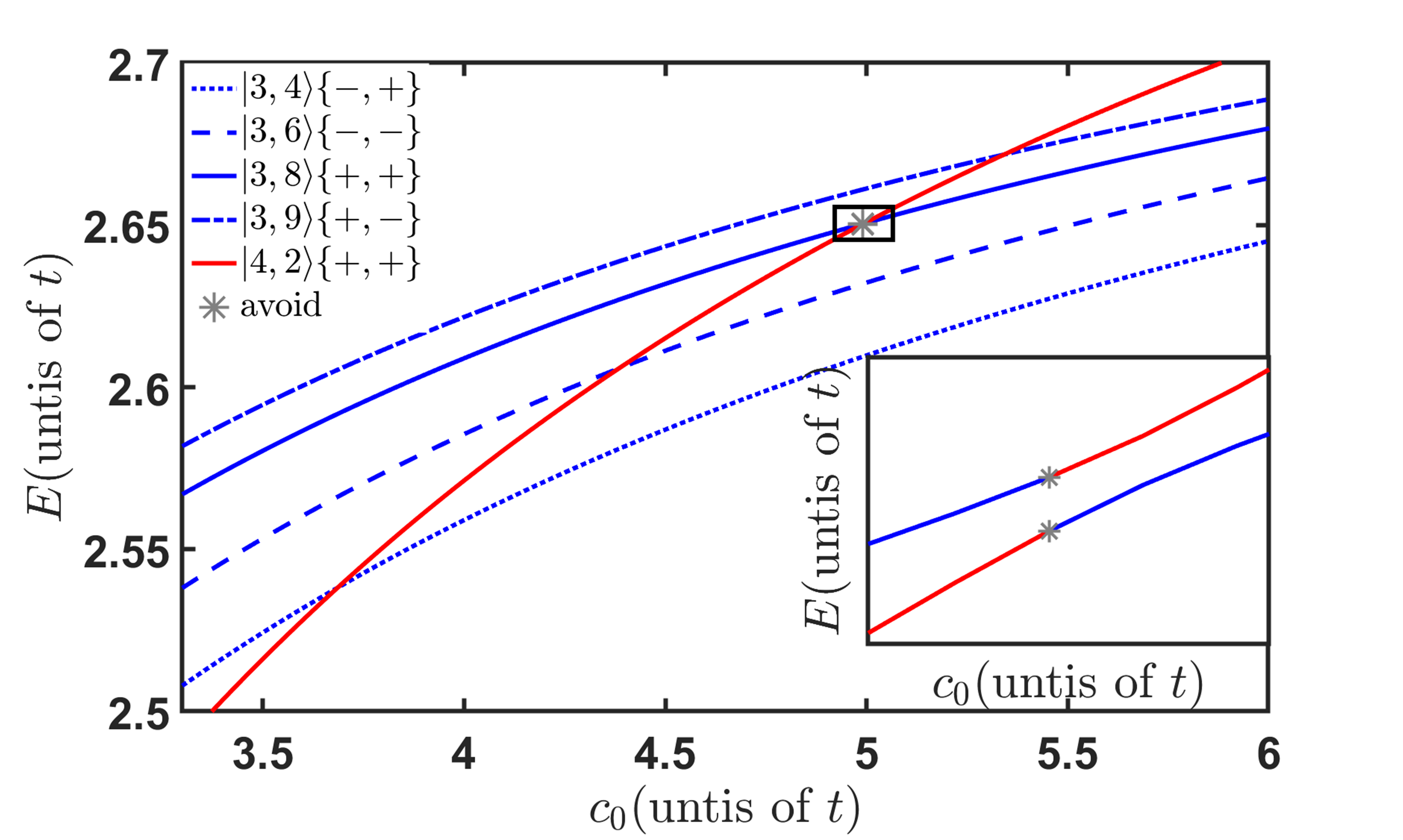}\caption{\label{fig-avoid} Crossings between $\left| {m = 3,k = {4,6,8,9}} \right\rangle $ and $\left| {m = 4,k = 2 }\right\rangle$,
of which the parities of $\{ \langle{{\hat {\cal R}}_{sp}\times{{\hat {\cal R}}_{ch}}\rangle}, \langle{{\hat {\cal S}}_{sp}}\rangle \}$ are explicitly
given. The inset zoom in the box of the main figure, presenting detailed view of avoided crossing. }
\end{figure}

The accidental degeneracy of eigenstates from different manifolds is also manifested as the crossing of the related
eigenenergies as a function of $c_0$, which can be further specified as the direct and/or avoided crossings. 
The symmetries of the eigenstates determine whether the crossing is a direct or an avoided one.
Figure \ref{fig-avoid} demonstrates the inter-manifold crossing between $\left| {m = 4,k = {2}}\right\rangle $ and  
$\left| {m = 3,k = {4,6,8,9}} \right\rangle$, and these eigenstates are of different parities with respect to 
${\hat {\cal S}}_{sp}$ and ${\hat {\cal R}}_{sp}\times{\hat {\cal R}}_{ch}$. 
It can be found that $\left| {m = 4,k = {2}}\right\rangle $ exhibits the direct crossing with $\left| {m = 3,k = {4,6,9}} \right\rangle$,
but the avoided crossing with $\left| {m = 3,k = 8 }\right\rangle$, as highlighted in the inset of the figure. 
The parities of these eigenstates with respect to ${\hat {\cal S}}_{sp}$ and ${\hat {\cal R}}_{sp}\times{\hat {\cal R}}_{ch}$
are given in the figure, and $\left| {m = 4,k = {2}}\right\rangle $ shares the same parities with  $\left| {m = 3,k = 8 }\right\rangle$,
but different from $\left| {m = 3,k = {4,6,9}} \right\rangle$.
It can then be deduced that the avoided crossing only takes place between eigenstates of the same parities with respect to
${\hat {\cal S}}_{sp}$ and ${\hat {\cal R}}_{sp}\times{\hat {\cal R}}_{ch}$.
This is attributed to the fact that the avoided crossing is induced by the coupling $\hat H_{sp-ch}$ between the spin and charge sectors,
and $\hat H_{sp-ch}$ commutes with ${\hat {\cal S}}_{sp}$ and ${\hat {\cal R}}_{sp}\times{\hat {\cal R}}_{ch}$, but not the individual 
${\hat {\cal R}}_{sp}$ or ${\hat {\cal R}}_{ch}$. $\hat H_{sp-ch}$ can then only couple  $|m\rangle_{ch}\times|k\rangle_{sp}$
of the same parities of the two symmetry operators. 
The avoided crossing between eigenstates of different manifolds is manifested as a signature of the coupling between
the spin and charge sectors in the intermediate interaction regime.

\section{Conclusion and Outlook}\label{IV}
In this work, we numerically investigate the spin-1 lattice gas in the intermediate interaction regime, 
which preserves the decomposition of the whole system into the spin and charge sectors as in the strong interaction regime.
In the intermediate interaction regime, however, energetical detuning between the two sectors is lowered and
the inter-sector coupling is enhanced, which is manifested as the major difference from the strong interaction regime.
The decomposition, for one thing, groups the low-lying eigenstates into different manifolds, 
and for another introduces the emergent spatial inversion symmetry separately in the spin and charge sectors to these eigenstates,
with hidden correlations. The lowered energetical detuning vanishes the inter-manifold gap in the eigenenergy
spectrum, and gives rise to the accidental degeneracy between eigenstates from different manifolds, which is also
manifested as the inter-manifold crossings.
The inter-sector coupling can induce the avoided crossings between accidentally degenerate eigenstates of the same
parity of the spin flipping and spatial inversion symmetry.

Our work focuses on the emergent spatial inversion symmetry and the inter-sector coupling effect on the low-lying
eigenstates of a finite spinor lattice gas. It would be interesting to extend the investigation to the dynamical
effects of the inter-sector coupling, since the eigenenergy spectrum already indicates that the higher excited states are
more strongly affected by the inter-sector coupling. Moreover, the coupling can be further enhanced by enlarging the size
of the system, which can vanish the mini-gaps between eigenstates, and the coupling effect would be not restricted 
to the accidentally degenerate eigenstates.



\begin{acknowledgments}
This work was supported by the Key Researh and Development Program of China 
(Grants No.2022YFA1404102, No. 2022YFC3003802 and No. 2021YFB3900204 ).
This work is also supported by the Cluster of  Excellence 'Advanced Imaging of Matter' of the Deutsche Forschungsgemeinschaft (DFG) - EXC 2056 - project ID 390715994.

\end{acknowledgments}

\bibliography{references}

\begin{thebibliography}{64}%
\makeatletter
\providecommand \@ifxundefined [1]{%
 \@ifx{#1\undefined}
}%
\providecommand \@ifnum [1]{%
 \ifnum #1\expandafter \@firstoftwo
 \else \expandafter \@secondoftwo
 \fi
}%
\providecommand \@ifx [1]{%
 \ifx #1\expandafter \@firstoftwo
 \else \expandafter \@secondoftwo
 \fi
}%
\providecommand \natexlab [1]{#1}%
\providecommand \enquote  [1]{``#1''}%
\providecommand \bibnamefont  [1]{#1}%
\providecommand \bibfnamefont [1]{#1}%
\providecommand \citenamefont [1]{#1}%
\providecommand \href@noop [0]{\@secondoftwo}%
\providecommand \href [0]{\begingroup \@sanitize@url \@href}%
\providecommand \@href[1]{\@@startlink{#1}\@@href}%
\providecommand \@@href[1]{\endgroup#1\@@endlink}%
\providecommand \@sanitize@url [0]{\catcode `\\12\catcode `\$12\catcode
  `\&12\catcode `\#12\catcode `\^12\catcode `\_12\catcode `\%12\relax}%
\providecommand \@@startlink[1]{}%
\providecommand \@@endlink[0]{}%
\providecommand \url  [0]{\begingroup\@sanitize@url \@url }%
\providecommand \@url [1]{\endgroup\@href {#1}{\urlprefix }}%
\providecommand \urlprefix  [0]{URL }%
\providecommand \Eprint [0]{\href }%
\providecommand \doibase [0]{https://doi.org/}%
\providecommand \selectlanguage [0]{\@gobble}%
\providecommand \bibinfo  [0]{\@secondoftwo}%
\providecommand \bibfield  [0]{\@secondoftwo}%
\providecommand \translation [1]{[#1]}%
\providecommand \BibitemOpen [0]{}%
\providecommand \bibitemStop [0]{}%
\providecommand \bibitemNoStop [0]{.\EOS\space}%
\providecommand \EOS [0]{\spacefactor3000\relax}%
\providecommand \BibitemShut  [1]{\csname bibitem#1\endcsname}%
\let\auto@bib@innerbib\@empty
\bibitem [{\citenamefont {Davis}\ \emph {et~al.}(1995)\citenamefont {Davis},
  \citenamefont {Mewes}, \citenamefont {Andrews}, \citenamefont {van Druten},
  \citenamefont {Durfee}, \citenamefont {Kurn},\ and\ \citenamefont
  {Ketterle}}]{Davis1995}%
  \BibitemOpen
  \bibfield  {author} {\bibinfo {author} {\bibfnamefont {K.~B.}\ \bibnamefont
  {Davis}}, \bibinfo {author} {\bibfnamefont {M.~O.}\ \bibnamefont {Mewes}},
  \bibinfo {author} {\bibfnamefont {M.~R.}\ \bibnamefont {Andrews}}, \bibinfo
  {author} {\bibfnamefont {N.~J.}\ \bibnamefont {van Druten}}, \bibinfo
  {author} {\bibfnamefont {D.~S.}\ \bibnamefont {Durfee}}, \bibinfo {author}
  {\bibfnamefont {D.~M.}\ \bibnamefont {Kurn}},\ and\ \bibinfo {author}
  {\bibfnamefont {W.}~\bibnamefont {Ketterle}},\ }\href
  {https://doi.org/10.1103/PhysRevLett.75.3969} {\bibfield  {journal} {\bibinfo
   {journal} {Phys. Rev. Lett.}\ }\textbf {\bibinfo {volume} {75}},\ \bibinfo
  {pages} {3969} (\bibinfo {year} {1995})}\BibitemShut {NoStop}%
\bibitem [{\citenamefont {Ketterle}(2002)}]{Ketterle2002}%
  \BibitemOpen
  \bibfield  {author} {\bibinfo {author} {\bibfnamefont {W.}~\bibnamefont
  {Ketterle}},\ }\href {https://doi.org/10.1103/RevModPhys.74.1131} {\bibfield
  {journal} {\bibinfo  {journal} {Rev. Mod. Phys.}\ }\textbf {\bibinfo {volume}
  {74}},\ \bibinfo {pages} {1131} (\bibinfo {year} {2002})}\BibitemShut
  {NoStop}%
\bibitem [{\citenamefont {Morsch}\ and\ \citenamefont
  {Oberthaler}(2006)}]{Morsch2006}%
  \BibitemOpen
  \bibfield  {author} {\bibinfo {author} {\bibfnamefont {O.}~\bibnamefont
  {Morsch}}\ and\ \bibinfo {author} {\bibfnamefont {M.}~\bibnamefont
  {Oberthaler}},\ }\href {https://doi.org/10.1103/RevModPhys.78.179} {\bibfield
   {journal} {\bibinfo  {journal} {Rev. Mod. Phys.}\ }\textbf {\bibinfo
  {volume} {78}},\ \bibinfo {pages} {179} (\bibinfo {year} {2006})}\BibitemShut
  {NoStop}%
\bibitem [{\citenamefont {Becker}\ \emph {et~al.}(2008)\citenamefont {Becker},
  \citenamefont {Stellmer}, \citenamefont {Soltan-Panahi}, \citenamefont
  {D{\"o}rscher}, \citenamefont {Baumert}, \citenamefont {Richter},
  \citenamefont {Kronj{\"a}ger}, \citenamefont {Bongs},\ and\ \citenamefont
  {Sengstock}}]{Becker2008}%
  \BibitemOpen
  \bibfield  {author} {\bibinfo {author} {\bibfnamefont {C.}~\bibnamefont
  {Becker}}, \bibinfo {author} {\bibfnamefont {S.}~\bibnamefont {Stellmer}},
  \bibinfo {author} {\bibfnamefont {P.}~\bibnamefont {Soltan-Panahi}}, \bibinfo
  {author} {\bibfnamefont {S.}~\bibnamefont {D{\"o}rscher}}, \bibinfo {author}
  {\bibfnamefont {M.}~\bibnamefont {Baumert}}, \bibinfo {author} {\bibfnamefont
  {E.-M.}\ \bibnamefont {Richter}}, \bibinfo {author} {\bibfnamefont
  {J.}~\bibnamefont {Kronj{\"a}ger}}, \bibinfo {author} {\bibfnamefont
  {K.}~\bibnamefont {Bongs}},\ and\ \bibinfo {author} {\bibfnamefont
  {K.}~\bibnamefont {Sengstock}},\ }\href
  {https://doi.org/https://doi.org/10.1038/nphys962} {\bibfield  {journal}
  {\bibinfo  {journal} {Nature Physics}\ }\textbf {\bibinfo {volume} {4}},\
  \bibinfo {pages} {496} (\bibinfo {year} {2008})}\BibitemShut {NoStop}%
\bibitem [{\citenamefont {Bloch}\ \emph {et~al.}(2008)\citenamefont {Bloch},
  \citenamefont {Dalibard},\ and\ \citenamefont {Zwerger}}]{Bloch2008}%
  \BibitemOpen
  \bibfield  {author} {\bibinfo {author} {\bibfnamefont {I.}~\bibnamefont
  {Bloch}}, \bibinfo {author} {\bibfnamefont {J.}~\bibnamefont {Dalibard}},\
  and\ \bibinfo {author} {\bibfnamefont {W.}~\bibnamefont {Zwerger}},\ }\href
  {https://doi.org/10.1103/RevModPhys.80.885} {\bibfield  {journal} {\bibinfo
  {journal} {Rev. Mod. Phys.}\ }\textbf {\bibinfo {volume} {80}},\ \bibinfo
  {pages} {885} (\bibinfo {year} {2008})}\BibitemShut {NoStop}%
\bibitem [{\citenamefont {DeMarco}\ and\ \citenamefont
  {Jin}(1999)}]{DeMarco1999}%
  \BibitemOpen
  \bibfield  {author} {\bibinfo {author} {\bibfnamefont {B.}~\bibnamefont
  {DeMarco}}\ and\ \bibinfo {author} {\bibfnamefont {D.~S.}\ \bibnamefont
  {Jin}},\ }\href {https://doi.org/10.1126/science.285.5434.1703} {\bibfield
  {journal} {\bibinfo  {journal} {Science}\ }\textbf {\bibinfo {volume}
  {285}},\ \bibinfo {pages} {1703} (\bibinfo {year} {1999})}\BibitemShut
  {NoStop}%
\bibitem [{\citenamefont {Zwierlein}\ \emph {et~al.}(2005)\citenamefont
  {Zwierlein}, \citenamefont {Abo-Shaeer}, \citenamefont {Schirotzek},
  \citenamefont {Schunck},\ and\ \citenamefont {Ketterle}}]{Zwierlein2005}%
  \BibitemOpen
  \bibfield  {author} {\bibinfo {author} {\bibfnamefont {M.~W.}\ \bibnamefont
  {Zwierlein}}, \bibinfo {author} {\bibfnamefont {J.~R.}\ \bibnamefont
  {Abo-Shaeer}}, \bibinfo {author} {\bibfnamefont {A.}~\bibnamefont
  {Schirotzek}}, \bibinfo {author} {\bibfnamefont {C.~H.}\ \bibnamefont
  {Schunck}},\ and\ \bibinfo {author} {\bibfnamefont {W.}~\bibnamefont
  {Ketterle}},\ }\href {https://doi.org/https://doi.org/10.1038/nature03858}
  {\bibfield  {journal} {\bibinfo  {journal} {Nature}\ }\textbf {\bibinfo
  {volume} {435}},\ \bibinfo {pages} {1047} (\bibinfo {year}
  {2005})}\BibitemShut {NoStop}%
\bibitem [{\citenamefont {Giorgini}\ \emph {et~al.}(2008)\citenamefont
  {Giorgini}, \citenamefont {Pitaevskii},\ and\ \citenamefont
  {Stringari}}]{Giorgini2008}%
  \BibitemOpen
  \bibfield  {author} {\bibinfo {author} {\bibfnamefont {S.}~\bibnamefont
  {Giorgini}}, \bibinfo {author} {\bibfnamefont {L.~P.}\ \bibnamefont
  {Pitaevskii}},\ and\ \bibinfo {author} {\bibfnamefont {S.}~\bibnamefont
  {Stringari}},\ }\href {https://doi.org/10.1103/RevModPhys.80.1215} {\bibfield
   {journal} {\bibinfo  {journal} {Rev. Mod. Phys.}\ }\textbf {\bibinfo
  {volume} {80}},\ \bibinfo {pages} {1215} (\bibinfo {year}
  {2008})}\BibitemShut {NoStop}%
\bibitem [{\citenamefont {Murmann}\ \emph {et~al.}(2015)\citenamefont
  {Murmann}, \citenamefont {Bergschneider}, \citenamefont {Klinkhamer},
  \citenamefont {Z\"urn}, \citenamefont {Lompe},\ and\ \citenamefont
  {Jochim}}]{Murmann2015}%
  \BibitemOpen
  \bibfield  {author} {\bibinfo {author} {\bibfnamefont {S.}~\bibnamefont
  {Murmann}}, \bibinfo {author} {\bibfnamefont {A.}~\bibnamefont
  {Bergschneider}}, \bibinfo {author} {\bibfnamefont {V.~M.}\ \bibnamefont
  {Klinkhamer}}, \bibinfo {author} {\bibfnamefont {G.}~\bibnamefont {Z\"urn}},
  \bibinfo {author} {\bibfnamefont {T.}~\bibnamefont {Lompe}},\ and\ \bibinfo
  {author} {\bibfnamefont {S.}~\bibnamefont {Jochim}},\ }\href
  {https://doi.org/10.1103/PhysRevLett.114.080402} {\bibfield  {journal}
  {\bibinfo  {journal} {Phys. Rev. Lett.}\ }\textbf {\bibinfo {volume} {114}},\
  \bibinfo {pages} {080402} (\bibinfo {year} {2015})}\BibitemShut {NoStop}%
\bibitem [{\citenamefont {Omran}\ \emph {et~al.}(2015)\citenamefont {Omran},
  \citenamefont {Boll}, \citenamefont {Hilker}, \citenamefont {Kleinlein},
  \citenamefont {Salomon}, \citenamefont {Bloch},\ and\ \citenamefont
  {Gross}}]{Omran2015}%
  \BibitemOpen
  \bibfield  {author} {\bibinfo {author} {\bibfnamefont {A.}~\bibnamefont
  {Omran}}, \bibinfo {author} {\bibfnamefont {M.}~\bibnamefont {Boll}},
  \bibinfo {author} {\bibfnamefont {T.~A.}\ \bibnamefont {Hilker}}, \bibinfo
  {author} {\bibfnamefont {K.}~\bibnamefont {Kleinlein}}, \bibinfo {author}
  {\bibfnamefont {G.}~\bibnamefont {Salomon}}, \bibinfo {author} {\bibfnamefont
  {I.}~\bibnamefont {Bloch}},\ and\ \bibinfo {author} {\bibfnamefont
  {C.}~\bibnamefont {Gross}},\ }\href
  {https://doi.org/10.1103/PhysRevLett.115.263001} {\bibfield  {journal}
  {\bibinfo  {journal} {Phys. Rev. Lett.}\ }\textbf {\bibinfo {volume} {115}},\
  \bibinfo {pages} {263001} (\bibinfo {year} {2015})}\BibitemShut {NoStop}%
\bibitem [{\citenamefont {Greiner}\ \emph {et~al.}(2002)\citenamefont
  {Greiner}, \citenamefont {Mandel}, \citenamefont {Esslinger}, \citenamefont
  {H{\"a}nsch},\ and\ \citenamefont {Bloch}}]{Greiner2002}%
  \BibitemOpen
  \bibfield  {author} {\bibinfo {author} {\bibfnamefont {M.}~\bibnamefont
  {Greiner}}, \bibinfo {author} {\bibfnamefont {O.}~\bibnamefont {Mandel}},
  \bibinfo {author} {\bibfnamefont {T.}~\bibnamefont {Esslinger}}, \bibinfo
  {author} {\bibfnamefont {T.~W.}\ \bibnamefont {H{\"a}nsch}},\ and\ \bibinfo
  {author} {\bibfnamefont {I.}~\bibnamefont {Bloch}},\ }\href
  {https://doi.org/https://doi.org/10.1038/415039a} {\bibfield  {journal}
  {\bibinfo  {journal} {nature}\ }\textbf {\bibinfo {volume} {415}},\ \bibinfo
  {pages} {39} (\bibinfo {year} {2002})}\BibitemShut {NoStop}%
\bibitem [{\citenamefont {Bloch}(2005)}]{Bloch2005}%
  \BibitemOpen
  \bibfield  {author} {\bibinfo {author} {\bibfnamefont {I.}~\bibnamefont
  {Bloch}},\ }\href {https://doi.org/https://doi.org/10.1038/nphys138}
  {\bibfield  {journal} {\bibinfo  {journal} {Nature physics}\ }\textbf
  {\bibinfo {volume} {1}},\ \bibinfo {pages} {23} (\bibinfo {year}
  {2005})}\BibitemShut {NoStop}%
\bibitem [{\citenamefont {Sidorenkov}\ \emph {et~al.}(2013)\citenamefont
  {Sidorenkov}, \citenamefont {Tey}, \citenamefont {Grimm}, \citenamefont
  {Hou}, \citenamefont {Pitaevskii},\ and\ \citenamefont
  {Stringari}}]{Sidorenkov2013}%
  \BibitemOpen
  \bibfield  {author} {\bibinfo {author} {\bibfnamefont {L.~A.}\ \bibnamefont
  {Sidorenkov}}, \bibinfo {author} {\bibfnamefont {M.~K.}\ \bibnamefont {Tey}},
  \bibinfo {author} {\bibfnamefont {R.}~\bibnamefont {Grimm}}, \bibinfo
  {author} {\bibfnamefont {Y.-H.}\ \bibnamefont {Hou}}, \bibinfo {author}
  {\bibfnamefont {L.}~\bibnamefont {Pitaevskii}},\ and\ \bibinfo {author}
  {\bibfnamefont {S.}~\bibnamefont {Stringari}},\ }\href
  {https://doi.org/https://doi.org/10.1038/nature12136} {\bibfield  {journal}
  {\bibinfo  {journal} {Nature}\ }\textbf {\bibinfo {volume} {498}},\ \bibinfo
  {pages} {78} (\bibinfo {year} {2013})}\BibitemShut {NoStop}%
\bibitem [{\citenamefont {Bernien}\ \emph {et~al.}(2017)\citenamefont
  {Bernien}, \citenamefont {Schwartz}, \citenamefont {Keesling}, \citenamefont
  {Levine}, \citenamefont {Omran}, \citenamefont {Pichler}, \citenamefont
  {Choi}, \citenamefont {Zibrov}, \citenamefont {Endres}, \citenamefont
  {Greiner} \emph {et~al.}}]{Bernien2017}%
  \BibitemOpen
  \bibfield  {author} {\bibinfo {author} {\bibfnamefont {H.}~\bibnamefont
  {Bernien}}, \bibinfo {author} {\bibfnamefont {S.}~\bibnamefont {Schwartz}},
  \bibinfo {author} {\bibfnamefont {A.}~\bibnamefont {Keesling}}, \bibinfo
  {author} {\bibfnamefont {H.}~\bibnamefont {Levine}}, \bibinfo {author}
  {\bibfnamefont {A.}~\bibnamefont {Omran}}, \bibinfo {author} {\bibfnamefont
  {H.}~\bibnamefont {Pichler}}, \bibinfo {author} {\bibfnamefont
  {S.}~\bibnamefont {Choi}}, \bibinfo {author} {\bibfnamefont {A.~S.}\
  \bibnamefont {Zibrov}}, \bibinfo {author} {\bibfnamefont {M.}~\bibnamefont
  {Endres}}, \bibinfo {author} {\bibfnamefont {M.}~\bibnamefont {Greiner}},
  \emph {et~al.},\ }\href {https://doi.org/https://doi.org/10.1038/nature24622}
  {\bibfield  {journal} {\bibinfo  {journal} {Nature}\ }\textbf {\bibinfo
  {volume} {551}},\ \bibinfo {pages} {579} (\bibinfo {year}
  {2017})}\BibitemShut {NoStop}%
\bibitem [{\citenamefont {Gross}\ and\ \citenamefont
  {Bloch}(2017)}]{Gross2017}%
  \BibitemOpen
  \bibfield  {author} {\bibinfo {author} {\bibfnamefont {C.}~\bibnamefont
  {Gross}}\ and\ \bibinfo {author} {\bibfnamefont {I.}~\bibnamefont {Bloch}},\
  }\href {https://doi.org/10.1126/science.aal3837} {\bibfield  {journal}
  {\bibinfo  {journal} {Science}\ }\textbf {\bibinfo {volume} {357}},\ \bibinfo
  {pages} {995} (\bibinfo {year} {2017})}\BibitemShut {NoStop}%
\bibitem [{\citenamefont {Chiu}\ \emph {et~al.}(2019)\citenamefont {Chiu},
  \citenamefont {Ji}, \citenamefont {Bohrdt}, \citenamefont {Xu}, \citenamefont
  {Knap}, \citenamefont {Demler}, \citenamefont {Grusdt}, \citenamefont
  {Greiner},\ and\ \citenamefont {Greif}}]{Chiu2019}%
  \BibitemOpen
  \bibfield  {author} {\bibinfo {author} {\bibfnamefont {C.~S.}\ \bibnamefont
  {Chiu}}, \bibinfo {author} {\bibfnamefont {G.}~\bibnamefont {Ji}}, \bibinfo
  {author} {\bibfnamefont {A.}~\bibnamefont {Bohrdt}}, \bibinfo {author}
  {\bibfnamefont {M.}~\bibnamefont {Xu}}, \bibinfo {author} {\bibfnamefont
  {M.}~\bibnamefont {Knap}}, \bibinfo {author} {\bibfnamefont {E.}~\bibnamefont
  {Demler}}, \bibinfo {author} {\bibfnamefont {F.}~\bibnamefont {Grusdt}},
  \bibinfo {author} {\bibfnamefont {M.}~\bibnamefont {Greiner}},\ and\ \bibinfo
  {author} {\bibfnamefont {D.}~\bibnamefont {Greif}},\ }\href
  {https://doi.org/10.1126/science.aav3587} {\bibfield  {journal} {\bibinfo
  {journal} {Science}\ }\textbf {\bibinfo {volume} {365}},\ \bibinfo {pages}
  {251} (\bibinfo {year} {2019})}\BibitemShut {NoStop}%
\bibitem [{\citenamefont {Sun}\ \emph {et~al.}(2018)\citenamefont {Sun},
  \citenamefont {Wang}, \citenamefont {Xu}, \citenamefont {Yi}, \citenamefont
  {Zhang}, \citenamefont {Wu}, \citenamefont {Deng}, \citenamefont {Liu},
  \citenamefont {Chen},\ and\ \citenamefont {Pan}}]{Sun2018}%
  \BibitemOpen
  \bibfield  {author} {\bibinfo {author} {\bibfnamefont {W.}~\bibnamefont
  {Sun}}, \bibinfo {author} {\bibfnamefont {B.-Z.}\ \bibnamefont {Wang}},
  \bibinfo {author} {\bibfnamefont {X.-T.}\ \bibnamefont {Xu}}, \bibinfo
  {author} {\bibfnamefont {C.-R.}\ \bibnamefont {Yi}}, \bibinfo {author}
  {\bibfnamefont {L.}~\bibnamefont {Zhang}}, \bibinfo {author} {\bibfnamefont
  {Z.}~\bibnamefont {Wu}}, \bibinfo {author} {\bibfnamefont {Y.}~\bibnamefont
  {Deng}}, \bibinfo {author} {\bibfnamefont {X.-J.}\ \bibnamefont {Liu}},
  \bibinfo {author} {\bibfnamefont {S.}~\bibnamefont {Chen}},\ and\ \bibinfo
  {author} {\bibfnamefont {J.-W.}\ \bibnamefont {Pan}},\ }\href
  {https://doi.org/10.1103/PhysRevLett.121.150401} {\bibfield  {journal}
  {\bibinfo  {journal} {Phys. Rev. Lett.}\ }\textbf {\bibinfo {volume} {121}},\
  \bibinfo {pages} {150401} (\bibinfo {year} {2018})}\BibitemShut {NoStop}%
\bibitem [{\citenamefont {Wang}\ \emph {et~al.}(2021)\citenamefont {Wang},
  \citenamefont {Cheng}, \citenamefont {Wang}, \citenamefont {Zhang},
  \citenamefont {Lu}, \citenamefont {Yi}, \citenamefont {Niu}, \citenamefont
  {Deng}, \citenamefont {Liu}, \citenamefont {Chen},\ and\ \citenamefont
  {Pan}}]{Wang2021}%
  \BibitemOpen
  \bibfield  {author} {\bibinfo {author} {\bibfnamefont {Z.-Y.}\ \bibnamefont
  {Wang}}, \bibinfo {author} {\bibfnamefont {X.-C.}\ \bibnamefont {Cheng}},
  \bibinfo {author} {\bibfnamefont {B.-Z.}\ \bibnamefont {Wang}}, \bibinfo
  {author} {\bibfnamefont {J.-Y.}\ \bibnamefont {Zhang}}, \bibinfo {author}
  {\bibfnamefont {Y.-H.}\ \bibnamefont {Lu}}, \bibinfo {author} {\bibfnamefont
  {C.-R.}\ \bibnamefont {Yi}}, \bibinfo {author} {\bibfnamefont
  {S.}~\bibnamefont {Niu}}, \bibinfo {author} {\bibfnamefont {Y.}~\bibnamefont
  {Deng}}, \bibinfo {author} {\bibfnamefont {X.-J.}\ \bibnamefont {Liu}},
  \bibinfo {author} {\bibfnamefont {S.}~\bibnamefont {Chen}},\ and\ \bibinfo
  {author} {\bibfnamefont {J.-W.}\ \bibnamefont {Pan}},\ }\href
  {https://doi.org/10.1126/science.abc0105} {\bibfield  {journal} {\bibinfo
  {journal} {Science}\ }\textbf {\bibinfo {volume} {372}},\ \bibinfo {pages}
  {271} (\bibinfo {year} {2021})}\BibitemShut {NoStop}%
\bibitem [{\citenamefont {Zhou}\ \emph {et~al.}(2022)\citenamefont {Zhou},
  \citenamefont {Su}, \citenamefont {Halimeh}, \citenamefont {Ott},
  \citenamefont {Sun}, \citenamefont {Hauke}, \citenamefont {Yang},
  \citenamefont {Yuan}, \citenamefont {Berges},\ and\ \citenamefont
  {Pan}}]{Zhou2022}%
  \BibitemOpen
  \bibfield  {author} {\bibinfo {author} {\bibfnamefont {Z.-Y.}\ \bibnamefont
  {Zhou}}, \bibinfo {author} {\bibfnamefont {G.-X.}\ \bibnamefont {Su}},
  \bibinfo {author} {\bibfnamefont {J.~C.}\ \bibnamefont {Halimeh}}, \bibinfo
  {author} {\bibfnamefont {R.}~\bibnamefont {Ott}}, \bibinfo {author}
  {\bibfnamefont {H.}~\bibnamefont {Sun}}, \bibinfo {author} {\bibfnamefont
  {P.}~\bibnamefont {Hauke}}, \bibinfo {author} {\bibfnamefont
  {B.}~\bibnamefont {Yang}}, \bibinfo {author} {\bibfnamefont {Z.-S.}\
  \bibnamefont {Yuan}}, \bibinfo {author} {\bibfnamefont {J.}~\bibnamefont
  {Berges}},\ and\ \bibinfo {author} {\bibfnamefont {J.-W.}\ \bibnamefont
  {Pan}},\ }\href {https://doi.org/10.1126/science.abl6277} {\bibfield
  {journal} {\bibinfo  {journal} {Science}\ }\textbf {\bibinfo {volume}
  {377}},\ \bibinfo {pages} {311} (\bibinfo {year} {2022})}\BibitemShut
  {NoStop}%
\bibitem [{\citenamefont {Duan}\ \emph {et~al.}(2003)\citenamefont {Duan},
  \citenamefont {Demler},\ and\ \citenamefont {Lukin}}]{Duan2003}%
  \BibitemOpen
  \bibfield  {author} {\bibinfo {author} {\bibfnamefont {L.-M.}\ \bibnamefont
  {Duan}}, \bibinfo {author} {\bibfnamefont {E.}~\bibnamefont {Demler}},\ and\
  \bibinfo {author} {\bibfnamefont {M.~D.}\ \bibnamefont {Lukin}},\ }\href
  {https://doi.org/10.1103/PhysRevLett.91.090402} {\bibfield  {journal}
  {\bibinfo  {journal} {Phys. Rev. Lett.}\ }\textbf {\bibinfo {volume} {91}},\
  \bibinfo {pages} {090402} (\bibinfo {year} {2003})}\BibitemShut {NoStop}%
\bibitem [{\citenamefont {Schneider}\ and\ \citenamefont
  {Saenz}(2012)}]{Schneider2012}%
  \BibitemOpen
  \bibfield  {author} {\bibinfo {author} {\bibfnamefont {P.-I.}\ \bibnamefont
  {Schneider}}\ and\ \bibinfo {author} {\bibfnamefont {A.}~\bibnamefont
  {Saenz}},\ }\href {https://doi.org/10.1103/PhysRevA.85.050304} {\bibfield
  {journal} {\bibinfo  {journal} {Phys. Rev. A}\ }\textbf {\bibinfo {volume}
  {85}},\ \bibinfo {pages} {050304} (\bibinfo {year} {2012})}\BibitemShut
  {NoStop}%
\bibitem [{\citenamefont {Shui}\ \emph {et~al.}(2021)\citenamefont {Shui},
  \citenamefont {Jin}, \citenamefont {Li}, \citenamefont {Wei}, \citenamefont
  {Chen}, \citenamefont {Li},\ and\ \citenamefont {Zhou}}]{Shui2021}%
  \BibitemOpen
  \bibfield  {author} {\bibinfo {author} {\bibfnamefont {H.}~\bibnamefont
  {Shui}}, \bibinfo {author} {\bibfnamefont {S.}~\bibnamefont {Jin}}, \bibinfo
  {author} {\bibfnamefont {Z.}~\bibnamefont {Li}}, \bibinfo {author}
  {\bibfnamefont {F.}~\bibnamefont {Wei}}, \bibinfo {author} {\bibfnamefont
  {X.}~\bibnamefont {Chen}}, \bibinfo {author} {\bibfnamefont {X.}~\bibnamefont
  {Li}},\ and\ \bibinfo {author} {\bibfnamefont {X.}~\bibnamefont {Zhou}},\
  }\href {https://doi.org/10.1103/PhysRevA.104.L060601} {\bibfield  {journal}
  {\bibinfo  {journal} {Phys. Rev. A}\ }\textbf {\bibinfo {volume} {104}},\
  \bibinfo {pages} {L060601} (\bibinfo {year} {2021})}\BibitemShut {NoStop}%
\bibitem [{\citenamefont {Graham}\ \emph {et~al.}(2022)\citenamefont {Graham},
  \citenamefont {Song}, \citenamefont {Scott}, \citenamefont {Poole},
  \citenamefont {Phuttitarn}, \citenamefont {Jooya}, \citenamefont {Eichler},
  \citenamefont {Jiang}, \citenamefont {Marra}, \citenamefont {Grinkemeyer}
  \emph {et~al.}}]{Graham2022}%
  \BibitemOpen
  \bibfield  {author} {\bibinfo {author} {\bibfnamefont {T.}~\bibnamefont
  {Graham}}, \bibinfo {author} {\bibfnamefont {Y.}~\bibnamefont {Song}},
  \bibinfo {author} {\bibfnamefont {J.}~\bibnamefont {Scott}}, \bibinfo
  {author} {\bibfnamefont {C.}~\bibnamefont {Poole}}, \bibinfo {author}
  {\bibfnamefont {L.}~\bibnamefont {Phuttitarn}}, \bibinfo {author}
  {\bibfnamefont {K.}~\bibnamefont {Jooya}}, \bibinfo {author} {\bibfnamefont
  {P.}~\bibnamefont {Eichler}}, \bibinfo {author} {\bibfnamefont
  {X.}~\bibnamefont {Jiang}}, \bibinfo {author} {\bibfnamefont
  {A.}~\bibnamefont {Marra}}, \bibinfo {author} {\bibfnamefont
  {B.}~\bibnamefont {Grinkemeyer}}, \emph {et~al.},\ }\href
  {https://doi.org/https://doi.org/10.1038/s41586-022-04603-6} {\bibfield
  {journal} {\bibinfo  {journal} {Nature}\ }\textbf {\bibinfo {volume} {604}},\
  \bibinfo {pages} {457} (\bibinfo {year} {2022})}\BibitemShut {NoStop}%
\bibitem [{\citenamefont {McDonnell}\ \emph {et~al.}(2022)\citenamefont
  {McDonnell}, \citenamefont {Keary},\ and\ \citenamefont
  {Pritchard}}]{McDonnell2022}%
  \BibitemOpen
  \bibfield  {author} {\bibinfo {author} {\bibfnamefont {K.}~\bibnamefont
  {McDonnell}}, \bibinfo {author} {\bibfnamefont {L.~F.}\ \bibnamefont
  {Keary}},\ and\ \bibinfo {author} {\bibfnamefont {J.~D.}\ \bibnamefont
  {Pritchard}},\ }\href {https://doi.org/10.1103/PhysRevLett.129.200501}
  {\bibfield  {journal} {\bibinfo  {journal} {Phys. Rev. Lett.}\ }\textbf
  {\bibinfo {volume} {129}},\ \bibinfo {pages} {200501} (\bibinfo {year}
  {2022})}\BibitemShut {NoStop}%
\bibitem [{\citenamefont {Fang}\ \emph {et~al.}(2022)\citenamefont {Fang},
  \citenamefont {Dai}, \citenamefont {Xiang}, \citenamefont {Chen},
  \citenamefont {Li}, \citenamefont {Gao}, \citenamefont {Zhu}, \citenamefont
  {Deng}, \citenamefont {Cao},\ and\ \citenamefont {Hu}}]{Fang2022}%
  \BibitemOpen
  \bibfield  {author} {\bibinfo {author} {\bibfnamefont {X.-T.}\ \bibnamefont
  {Fang}}, \bibinfo {author} {\bibfnamefont {Z.-Q.}\ \bibnamefont {Dai}},
  \bibinfo {author} {\bibfnamefont {D.}~\bibnamefont {Xiang}}, \bibinfo
  {author} {\bibfnamefont {S.-L.}\ \bibnamefont {Chen}}, \bibinfo {author}
  {\bibfnamefont {S.-J.}\ \bibnamefont {Li}}, \bibinfo {author} {\bibfnamefont
  {X.}~\bibnamefont {Gao}}, \bibinfo {author} {\bibfnamefont {Q.-R.}\
  \bibnamefont {Zhu}}, \bibinfo {author} {\bibfnamefont {X.}~\bibnamefont
  {Deng}}, \bibinfo {author} {\bibfnamefont {L.}~\bibnamefont {Cao}},\ and\
  \bibinfo {author} {\bibfnamefont {Z.-K.}\ \bibnamefont {Hu}},\ }\href
  {https://doi.org/10.1103/PhysRevA.106.033315} {\bibfield  {journal} {\bibinfo
   {journal} {Phys. Rev. A}\ }\textbf {\bibinfo {volume} {106}},\ \bibinfo
  {pages} {033315} (\bibinfo {year} {2022})}\BibitemShut {NoStop}%
\bibitem [{\citenamefont {González-Cuadra}\ \emph {et~al.}(2023)\citenamefont
  {González-Cuadra}, \citenamefont {Bluvstein}, \citenamefont {Kalinowski},
  \citenamefont {Kaubruegger}, \citenamefont {Maskara}, \citenamefont
  {Naldesi}, \citenamefont {Zache}, \citenamefont {Kaufman}, \citenamefont
  {Lukin}, \citenamefont {Pichler}, \citenamefont {Vermersch}, \citenamefont
  {Ye},\ and\ \citenamefont {Zoller}}]{GonzalezCuadra2023}%
  \BibitemOpen
  \bibfield  {author} {\bibinfo {author} {\bibfnamefont {D.}~\bibnamefont
  {González-Cuadra}}, \bibinfo {author} {\bibfnamefont {D.}~\bibnamefont
  {Bluvstein}}, \bibinfo {author} {\bibfnamefont {M.}~\bibnamefont
  {Kalinowski}}, \bibinfo {author} {\bibfnamefont {R.}~\bibnamefont
  {Kaubruegger}}, \bibinfo {author} {\bibfnamefont {N.}~\bibnamefont
  {Maskara}}, \bibinfo {author} {\bibfnamefont {P.}~\bibnamefont {Naldesi}},
  \bibinfo {author} {\bibfnamefont {T.~V.}\ \bibnamefont {Zache}}, \bibinfo
  {author} {\bibfnamefont {A.~M.}\ \bibnamefont {Kaufman}}, \bibinfo {author}
  {\bibfnamefont {M.~D.}\ \bibnamefont {Lukin}}, \bibinfo {author}
  {\bibfnamefont {H.}~\bibnamefont {Pichler}}, \bibinfo {author} {\bibfnamefont
  {B.}~\bibnamefont {Vermersch}}, \bibinfo {author} {\bibfnamefont
  {J.}~\bibnamefont {Ye}},\ and\ \bibinfo {author} {\bibfnamefont
  {P.}~\bibnamefont {Zoller}},\ }\bibfield  {journal} {\bibinfo  {journal}
  {Proceedings of the National Academy of Sciences}\ }\textbf {\bibinfo
  {volume} {120}},\ \href {https://doi.org/10.1073/pnas.2304294120}
  {10.1073/pnas.2304294120} (\bibinfo {year} {2023})\BibitemShut {NoStop}%
\bibitem [{\citenamefont {Swallows}\ \emph {et~al.}(2011)\citenamefont
  {Swallows}, \citenamefont {Bishof}, \citenamefont {Lin}, \citenamefont
  {Blatt}, \citenamefont {Martin}, \citenamefont {Rey},\ and\ \citenamefont
  {Ye}}]{Swallows2011}%
  \BibitemOpen
  \bibfield  {author} {\bibinfo {author} {\bibfnamefont {M.~D.}\ \bibnamefont
  {Swallows}}, \bibinfo {author} {\bibfnamefont {M.}~\bibnamefont {Bishof}},
  \bibinfo {author} {\bibfnamefont {Y.}~\bibnamefont {Lin}}, \bibinfo {author}
  {\bibfnamefont {S.}~\bibnamefont {Blatt}}, \bibinfo {author} {\bibfnamefont
  {M.~J.}\ \bibnamefont {Martin}}, \bibinfo {author} {\bibfnamefont {A.~M.}\
  \bibnamefont {Rey}},\ and\ \bibinfo {author} {\bibfnamefont {J.}~\bibnamefont
  {Ye}},\ }\href {https://doi.org/10.1126/science.1196442} {\bibfield
  {journal} {\bibinfo  {journal} {Science}\ }\textbf {\bibinfo {volume}
  {331}},\ \bibinfo {pages} {1043} (\bibinfo {year} {2011})}\BibitemShut
  {NoStop}%
\bibitem [{\citenamefont {Pezz\`e}\ \emph {et~al.}(2018)\citenamefont
  {Pezz\`e}, \citenamefont {Smerzi}, \citenamefont {Oberthaler}, \citenamefont
  {Schmied},\ and\ \citenamefont {Treutlein}}]{Pezze2018}%
  \BibitemOpen
  \bibfield  {author} {\bibinfo {author} {\bibfnamefont {L.}~\bibnamefont
  {Pezz\`e}}, \bibinfo {author} {\bibfnamefont {A.}~\bibnamefont {Smerzi}},
  \bibinfo {author} {\bibfnamefont {M.~K.}\ \bibnamefont {Oberthaler}},
  \bibinfo {author} {\bibfnamefont {R.}~\bibnamefont {Schmied}},\ and\ \bibinfo
  {author} {\bibfnamefont {P.}~\bibnamefont {Treutlein}},\ }\href
  {https://doi.org/10.1103/RevModPhys.90.035005} {\bibfield  {journal}
  {\bibinfo  {journal} {Rev. Mod. Phys.}\ }\textbf {\bibinfo {volume} {90}},\
  \bibinfo {pages} {035005} (\bibinfo {year} {2018})}\BibitemShut {NoStop}%
\bibitem [{\citenamefont {Fra\"{\i}sse}\ \emph {et~al.}(2019)\citenamefont
  {Fra\"{\i}sse}, \citenamefont {Baak},\ and\ \citenamefont
  {Fischer}}]{Fraisse2019}%
  \BibitemOpen
  \bibfield  {author} {\bibinfo {author} {\bibfnamefont {J.~M.~E.}\
  \bibnamefont {Fra\"{\i}sse}}, \bibinfo {author} {\bibfnamefont {J.-G.}\
  \bibnamefont {Baak}},\ and\ \bibinfo {author} {\bibfnamefont {U.~R.}\
  \bibnamefont {Fischer}},\ }\href {https://doi.org/10.1103/PhysRevA.99.043618}
  {\bibfield  {journal} {\bibinfo  {journal} {Phys. Rev. A}\ }\textbf {\bibinfo
  {volume} {99}},\ \bibinfo {pages} {043618} (\bibinfo {year}
  {2019})}\BibitemShut {NoStop}%
\bibitem [{\citenamefont {Liu}\ \emph {et~al.}(2022)\citenamefont {Liu},
  \citenamefont {Wu}, \citenamefont {Cao}, \citenamefont {Mao}, \citenamefont
  {Li}, \citenamefont {Guo}, \citenamefont {Tey},\ and\ \citenamefont
  {You}}]{Liu2022}%
  \BibitemOpen
  \bibfield  {author} {\bibinfo {author} {\bibfnamefont {Q.}~\bibnamefont
  {Liu}}, \bibinfo {author} {\bibfnamefont {L.-N.}\ \bibnamefont {Wu}},
  \bibinfo {author} {\bibfnamefont {J.-H.}\ \bibnamefont {Cao}}, \bibinfo
  {author} {\bibfnamefont {T.-W.}\ \bibnamefont {Mao}}, \bibinfo {author}
  {\bibfnamefont {X.-W.}\ \bibnamefont {Li}}, \bibinfo {author} {\bibfnamefont
  {S.-F.}\ \bibnamefont {Guo}}, \bibinfo {author} {\bibfnamefont {M.~K.}\
  \bibnamefont {Tey}},\ and\ \bibinfo {author} {\bibfnamefont {L.}~\bibnamefont
  {You}},\ }\href {https://doi.org/https://doi.org/10.1038/s41567-021-01441-7}
  {\bibfield  {journal} {\bibinfo  {journal} {Nature Physics}\ }\textbf
  {\bibinfo {volume} {18}},\ \bibinfo {pages} {167} (\bibinfo {year}
  {2022})}\BibitemShut {NoStop}%
\bibitem [{\citenamefont {Aeppli}\ \emph {et~al.}(2024)\citenamefont {Aeppli},
  \citenamefont {Kim}, \citenamefont {Warfield}, \citenamefont {Safronova},\
  and\ \citenamefont {Ye}}]{Aeppli2024}%
  \BibitemOpen
  \bibfield  {author} {\bibinfo {author} {\bibfnamefont {A.}~\bibnamefont
  {Aeppli}}, \bibinfo {author} {\bibfnamefont {K.}~\bibnamefont {Kim}},
  \bibinfo {author} {\bibfnamefont {W.}~\bibnamefont {Warfield}}, \bibinfo
  {author} {\bibfnamefont {M.~S.}\ \bibnamefont {Safronova}},\ and\ \bibinfo
  {author} {\bibfnamefont {J.}~\bibnamefont {Ye}},\ }\href
  {https://doi.org/10.1103/PhysRevLett.133.023401} {\bibfield  {journal}
  {\bibinfo  {journal} {Phys. Rev. Lett.}\ }\textbf {\bibinfo {volume} {133}},\
  \bibinfo {pages} {023401} (\bibinfo {year} {2024})}\BibitemShut {NoStop}%
\bibitem [{\citenamefont {Deuretzbacher}\ \emph {et~al.}(2017)\citenamefont
  {Deuretzbacher}, \citenamefont {Becker}, \citenamefont {Bjerlin},
  \citenamefont {Reimann},\ and\ \citenamefont {Santos}}]{Deuretzbacher2017}%
  \BibitemOpen
  \bibfield  {author} {\bibinfo {author} {\bibfnamefont {F.}~\bibnamefont
  {Deuretzbacher}}, \bibinfo {author} {\bibfnamefont {D.}~\bibnamefont
  {Becker}}, \bibinfo {author} {\bibfnamefont {J.}~\bibnamefont {Bjerlin}},
  \bibinfo {author} {\bibfnamefont {S.~M.}\ \bibnamefont {Reimann}},\ and\
  \bibinfo {author} {\bibfnamefont {L.}~\bibnamefont {Santos}},\ }\href
  {https://doi.org/10.1103/PhysRevA.95.043630} {\bibfield  {journal} {\bibinfo
  {journal} {Phys. Rev. A}\ }\textbf {\bibinfo {volume} {95}},\ \bibinfo
  {pages} {043630} (\bibinfo {year} {2017})}\BibitemShut {NoStop}%
\bibitem [{\citenamefont {Chung}\ \emph {et~al.}(2021)\citenamefont {Chung},
  \citenamefont {de~Hond}, \citenamefont {Xiang}, \citenamefont
  {Cruz-Col\'on},\ and\ \citenamefont {Ketterle}}]{Chung2021}%
  \BibitemOpen
  \bibfield  {author} {\bibinfo {author} {\bibfnamefont {W.~C.}\ \bibnamefont
  {Chung}}, \bibinfo {author} {\bibfnamefont {J.}~\bibnamefont {de~Hond}},
  \bibinfo {author} {\bibfnamefont {J.}~\bibnamefont {Xiang}}, \bibinfo
  {author} {\bibfnamefont {E.}~\bibnamefont {Cruz-Col\'on}},\ and\ \bibinfo
  {author} {\bibfnamefont {W.}~\bibnamefont {Ketterle}},\ }\href
  {https://doi.org/10.1103/PhysRevLett.126.163203} {\bibfield  {journal}
  {\bibinfo  {journal} {Phys. Rev. Lett.}\ }\textbf {\bibinfo {volume} {126}},\
  \bibinfo {pages} {163203} (\bibinfo {year} {2021})}\BibitemShut {NoStop}%
\bibitem [{\citenamefont {Garc\'{\i}a-Ripoll}\ \emph
  {et~al.}(2004)\citenamefont {Garc\'{\i}a-Ripoll}, \citenamefont
  {Martin-Delgado},\ and\ \citenamefont {Cirac}}]{GarciaRipoll2004}%
  \BibitemOpen
  \bibfield  {author} {\bibinfo {author} {\bibfnamefont {J.~J.}\ \bibnamefont
  {Garc\'{\i}a-Ripoll}}, \bibinfo {author} {\bibfnamefont {M.~A.}\ \bibnamefont
  {Martin-Delgado}},\ and\ \bibinfo {author} {\bibfnamefont {J.~I.}\
  \bibnamefont {Cirac}},\ }\href
  {https://doi.org/10.1103/PhysRevLett.93.250405} {\bibfield  {journal}
  {\bibinfo  {journal} {Phys. Rev. Lett.}\ }\textbf {\bibinfo {volume} {93}},\
  \bibinfo {pages} {250405} (\bibinfo {year} {2004})}\BibitemShut {NoStop}%
\bibitem [{\citenamefont {Chung}\ and\ \citenamefont {Yip}(2009)}]{Chung2009}%
  \BibitemOpen
  \bibfield  {author} {\bibinfo {author} {\bibfnamefont {M.-C.}\ \bibnamefont
  {Chung}}\ and\ \bibinfo {author} {\bibfnamefont {S.}~\bibnamefont {Yip}},\
  }\href {https://doi.org/10.1103/PhysRevA.80.053615} {\bibfield  {journal}
  {\bibinfo  {journal} {Phys. Rev. A}\ }\textbf {\bibinfo {volume} {80}},\
  \bibinfo {pages} {053615} (\bibinfo {year} {2009})}\BibitemShut {NoStop}%
\bibitem [{\citenamefont {Zhu}\ and\ \citenamefont {Zhao}(2018)}]{Zhu2018}%
  \BibitemOpen
  \bibfield  {author} {\bibinfo {author} {\bibfnamefont {M.-J.}\ \bibnamefont
  {Zhu}}\ and\ \bibinfo {author} {\bibfnamefont {B.}~\bibnamefont {Zhao}},\
  }\href {https://doi.org/https://doi.org/10.1140/epjd/e2018-80609-x}
  {\bibfield  {journal} {\bibinfo  {journal} {The European Physical Journal D}\
  }\textbf {\bibinfo {volume} {72}},\ \bibinfo {pages} {1} (\bibinfo {year}
  {2018})}\BibitemShut {NoStop}%
\bibitem [{\citenamefont {Katsura}\ and\ \citenamefont
  {Tasaki}(2013)}]{Katsura2013}%
  \BibitemOpen
  \bibfield  {author} {\bibinfo {author} {\bibfnamefont {H.}~\bibnamefont
  {Katsura}}\ and\ \bibinfo {author} {\bibfnamefont {H.}~\bibnamefont
  {Tasaki}},\ }\href {https://doi.org/10.1103/PhysRevLett.110.130405}
  {\bibfield  {journal} {\bibinfo  {journal} {Phys. Rev. Lett.}\ }\textbf
  {\bibinfo {volume} {110}},\ \bibinfo {pages} {130405} (\bibinfo {year}
  {2013})}\BibitemShut {NoStop}%
\bibitem [{\citenamefont {Batrouni}\ \emph {et~al.}(2009)\citenamefont
  {Batrouni}, \citenamefont {Rousseau},\ and\ \citenamefont
  {Scalettar}}]{PhysRevLett.102.140402}%
  \BibitemOpen
  \bibfield  {author} {\bibinfo {author} {\bibfnamefont {G.~G.}\ \bibnamefont
  {Batrouni}}, \bibinfo {author} {\bibfnamefont {V.~G.}\ \bibnamefont
  {Rousseau}},\ and\ \bibinfo {author} {\bibfnamefont {R.~T.}\ \bibnamefont
  {Scalettar}},\ }\href {https://doi.org/10.1103/PhysRevLett.102.140402}
  {\bibfield  {journal} {\bibinfo  {journal} {Phys. Rev. Lett.}\ }\textbf
  {\bibinfo {volume} {102}},\ \bibinfo {pages} {140402} (\bibinfo {year}
  {2009})}\BibitemShut {NoStop}%
\bibitem [{\citenamefont {Yang}\ and\ \citenamefont
  {Katsura}(2019{\natexlab{a}})}]{PhysRevLett.122.053401}%
  \BibitemOpen
  \bibfield  {author} {\bibinfo {author} {\bibfnamefont {H.}~\bibnamefont
  {Yang}}\ and\ \bibinfo {author} {\bibfnamefont {H.}~\bibnamefont {Katsura}},\
  }\href {https://doi.org/10.1103/PhysRevLett.122.053401} {\bibfield  {journal}
  {\bibinfo  {journal} {Phys. Rev. Lett.}\ }\textbf {\bibinfo {volume} {122}},\
  \bibinfo {pages} {053401} (\bibinfo {year} {2019}{\natexlab{a}})}\BibitemShut
  {NoStop}%
\bibitem [{\citenamefont {de~Forges~de Parny}\ and\ \citenamefont
  {Rousseau}(2018)}]{PhysRevA.97.023628}%
  \BibitemOpen
  \bibfield  {author} {\bibinfo {author} {\bibfnamefont {L.}~\bibnamefont
  {de~Forges~de Parny}}\ and\ \bibinfo {author} {\bibfnamefont {V.~G.}\
  \bibnamefont {Rousseau}},\ }\href
  {https://doi.org/10.1103/PhysRevA.97.023628} {\bibfield  {journal} {\bibinfo
  {journal} {Phys. Rev. A}\ }\textbf {\bibinfo {volume} {97}},\ \bibinfo
  {pages} {023628} (\bibinfo {year} {2018})}\BibitemShut {NoStop}%
\bibitem [{\citenamefont {Zeiher}\ \emph {et~al.}(2017)\citenamefont {Zeiher},
  \citenamefont {Choi}, \citenamefont {Rubio-Abadal}, \citenamefont {Pohl},
  \citenamefont {van Bijnen}, \citenamefont {Bloch},\ and\ \citenamefont
  {Gross}}]{Zeiher2017}%
  \BibitemOpen
  \bibfield  {author} {\bibinfo {author} {\bibfnamefont {J.}~\bibnamefont
  {Zeiher}}, \bibinfo {author} {\bibfnamefont {J.-y.}\ \bibnamefont {Choi}},
  \bibinfo {author} {\bibfnamefont {A.}~\bibnamefont {Rubio-Abadal}}, \bibinfo
  {author} {\bibfnamefont {T.}~\bibnamefont {Pohl}}, \bibinfo {author}
  {\bibfnamefont {R.}~\bibnamefont {van Bijnen}}, \bibinfo {author}
  {\bibfnamefont {I.}~\bibnamefont {Bloch}},\ and\ \bibinfo {author}
  {\bibfnamefont {C.}~\bibnamefont {Gross}},\ }\href
  {https://doi.org/10.1103/PhysRevX.7.041063} {\bibfield  {journal} {\bibinfo
  {journal} {Phys. Rev. X}\ }\textbf {\bibinfo {volume} {7}},\ \bibinfo {pages}
  {041063} (\bibinfo {year} {2017})}\BibitemShut {NoStop}%
\bibitem [{\citenamefont {Sun}\ \emph {et~al.}(2021)\citenamefont {Sun},
  \citenamefont {Yang}, \citenamefont {Wang}, \citenamefont {Zhou},
  \citenamefont {Su}, \citenamefont {Dai}, \citenamefont {Yuan},\ and\
  \citenamefont {Pan}}]{Sun2021}%
  \BibitemOpen
  \bibfield  {author} {\bibinfo {author} {\bibfnamefont {H.}~\bibnamefont
  {Sun}}, \bibinfo {author} {\bibfnamefont {B.}~\bibnamefont {Yang}}, \bibinfo
  {author} {\bibfnamefont {H.-Y.}\ \bibnamefont {Wang}}, \bibinfo {author}
  {\bibfnamefont {Z.-Y.}\ \bibnamefont {Zhou}}, \bibinfo {author}
  {\bibfnamefont {G.-X.}\ \bibnamefont {Su}}, \bibinfo {author} {\bibfnamefont
  {H.-N.}\ \bibnamefont {Dai}}, \bibinfo {author} {\bibfnamefont {Z.-S.}\
  \bibnamefont {Yuan}},\ and\ \bibinfo {author} {\bibfnamefont {J.-W.}\
  \bibnamefont {Pan}},\ }\href
  {https://doi.org/https://doi.org/10.1038/s41567-021-01277-1} {\bibfield
  {journal} {\bibinfo  {journal} {Nature Physics}\ }\textbf {\bibinfo {volume}
  {17}},\ \bibinfo {pages} {990} (\bibinfo {year} {2021})}\BibitemShut
  {NoStop}%
\bibitem [{\citenamefont {Kleine}\ \emph {et~al.}(2008)\citenamefont {Kleine},
  \citenamefont {Kollath}, \citenamefont {McCulloch}, \citenamefont
  {Giamarchi},\ and\ \citenamefont {Schollw\"ock}}]{Kleine2008}%
  \BibitemOpen
  \bibfield  {author} {\bibinfo {author} {\bibfnamefont {A.}~\bibnamefont
  {Kleine}}, \bibinfo {author} {\bibfnamefont {C.}~\bibnamefont {Kollath}},
  \bibinfo {author} {\bibfnamefont {I.~P.}\ \bibnamefont {McCulloch}}, \bibinfo
  {author} {\bibfnamefont {T.}~\bibnamefont {Giamarchi}},\ and\ \bibinfo
  {author} {\bibfnamefont {U.}~\bibnamefont {Schollw\"ock}},\ }\href
  {https://doi.org/10.1103/PhysRevA.77.013607} {\bibfield  {journal} {\bibinfo
  {journal} {Phys. Rev. A}\ }\textbf {\bibinfo {volume} {77}},\ \bibinfo
  {pages} {013607} (\bibinfo {year} {2008})}\BibitemShut {NoStop}%
\bibitem [{\citenamefont {de~Hond}\ \emph {et~al.}(2022)\citenamefont
  {de~Hond}, \citenamefont {Xiang}, \citenamefont {Chung}, \citenamefont
  {Cruz-Col\'on}, \citenamefont {Chen}, \citenamefont {Burton}, \citenamefont
  {Kennedy},\ and\ \citenamefont {Ketterle}}]{Hond2022}%
  \BibitemOpen
  \bibfield  {author} {\bibinfo {author} {\bibfnamefont {J.}~\bibnamefont
  {de~Hond}}, \bibinfo {author} {\bibfnamefont {J.}~\bibnamefont {Xiang}},
  \bibinfo {author} {\bibfnamefont {W.~C.}\ \bibnamefont {Chung}}, \bibinfo
  {author} {\bibfnamefont {E.}~\bibnamefont {Cruz-Col\'on}}, \bibinfo {author}
  {\bibfnamefont {W.}~\bibnamefont {Chen}}, \bibinfo {author} {\bibfnamefont
  {W.~C.}\ \bibnamefont {Burton}}, \bibinfo {author} {\bibfnamefont {C.~J.}\
  \bibnamefont {Kennedy}},\ and\ \bibinfo {author} {\bibfnamefont
  {W.}~\bibnamefont {Ketterle}},\ }\href
  {https://doi.org/10.1103/PhysRevLett.128.093401} {\bibfield  {journal}
  {\bibinfo  {journal} {Phys. Rev. Lett.}\ }\textbf {\bibinfo {volume} {128}},\
  \bibinfo {pages} {093401} (\bibinfo {year} {2022})}\BibitemShut {NoStop}%
\bibitem [{\citenamefont {Yip}(2003)}]{Yip2003}%
  \BibitemOpen
  \bibfield  {author} {\bibinfo {author} {\bibfnamefont {S.~K.}\ \bibnamefont
  {Yip}},\ }\href {https://doi.org/10.1103/PhysRevLett.90.250402} {\bibfield
  {journal} {\bibinfo  {journal} {Phys. Rev. Lett.}\ }\textbf {\bibinfo
  {volume} {90}},\ \bibinfo {pages} {250402} (\bibinfo {year}
  {2003})}\BibitemShut {NoStop}%
\bibitem [{\citenamefont {Imambekov}\ \emph {et~al.}(2003)\citenamefont
  {Imambekov}, \citenamefont {Lukin},\ and\ \citenamefont
  {Demler}}]{Imambekov2003}%
  \BibitemOpen
  \bibfield  {author} {\bibinfo {author} {\bibfnamefont {A.}~\bibnamefont
  {Imambekov}}, \bibinfo {author} {\bibfnamefont {M.}~\bibnamefont {Lukin}},\
  and\ \bibinfo {author} {\bibfnamefont {E.}~\bibnamefont {Demler}},\ }\href
  {https://doi.org/10.1103/PhysRevA.68.063602} {\bibfield  {journal} {\bibinfo
  {journal} {Phys. Rev. A}\ }\textbf {\bibinfo {volume} {68}},\ \bibinfo
  {pages} {063602} (\bibinfo {year} {2003})}\BibitemShut {NoStop}%
\bibitem [{\citenamefont {Rizzi}\ \emph {et~al.}(2005)\citenamefont {Rizzi},
  \citenamefont {Rossini}, \citenamefont {De~Chiara}, \citenamefont
  {Montangero},\ and\ \citenamefont {Fazio}}]{Rizzi2005}%
  \BibitemOpen
  \bibfield  {author} {\bibinfo {author} {\bibfnamefont {M.}~\bibnamefont
  {Rizzi}}, \bibinfo {author} {\bibfnamefont {D.}~\bibnamefont {Rossini}},
  \bibinfo {author} {\bibfnamefont {G.}~\bibnamefont {De~Chiara}}, \bibinfo
  {author} {\bibfnamefont {S.}~\bibnamefont {Montangero}},\ and\ \bibinfo
  {author} {\bibfnamefont {R.}~\bibnamefont {Fazio}},\ }\href
  {https://doi.org/10.1103/PhysRevLett.95.240404} {\bibfield  {journal}
  {\bibinfo  {journal} {Phys. Rev. Lett.}\ }\textbf {\bibinfo {volume} {95}},\
  \bibinfo {pages} {240404} (\bibinfo {year} {2005})}\BibitemShut {NoStop}%
\bibitem [{\citenamefont {Guan}\ \emph {et~al.}(2008)\citenamefont {Guan},
  \citenamefont {Batchelor}, \citenamefont {Lee},\ and\ \citenamefont
  {Zhou}}]{Guan2008}%
  \BibitemOpen
  \bibfield  {author} {\bibinfo {author} {\bibfnamefont {X.~W.}\ \bibnamefont
  {Guan}}, \bibinfo {author} {\bibfnamefont {M.~T.}\ \bibnamefont {Batchelor}},
  \bibinfo {author} {\bibfnamefont {C.}~\bibnamefont {Lee}},\ and\ \bibinfo
  {author} {\bibfnamefont {H.-Q.}\ \bibnamefont {Zhou}},\ }\href
  {https://doi.org/10.1103/PhysRevLett.100.200401} {\bibfield  {journal}
  {\bibinfo  {journal} {Phys. Rev. Lett.}\ }\textbf {\bibinfo {volume} {100}},\
  \bibinfo {pages} {200401} (\bibinfo {year} {2008})}\BibitemShut {NoStop}%
\bibitem [{\citenamefont {Wu}\ \emph {et~al.}(2003)\citenamefont {Wu},
  \citenamefont {Hu},\ and\ \citenamefont {Zhang}}]{Wu2003}%
  \BibitemOpen
  \bibfield  {author} {\bibinfo {author} {\bibfnamefont {C.}~\bibnamefont
  {Wu}}, \bibinfo {author} {\bibfnamefont {J.-p.}\ \bibnamefont {Hu}},\ and\
  \bibinfo {author} {\bibfnamefont {S.-c.}\ \bibnamefont {Zhang}},\ }\href
  {https://doi.org/10.1103/PhysRevLett.91.186402} {\bibfield  {journal}
  {\bibinfo  {journal} {Phys. Rev. Lett.}\ }\textbf {\bibinfo {volume} {91}},\
  \bibinfo {pages} {186402} (\bibinfo {year} {2003})}\BibitemShut {NoStop}%
\bibitem [{\citenamefont {WU}(2006)}]{WU2006}%
  \BibitemOpen
  \bibfield  {author} {\bibinfo {author} {\bibfnamefont {C.}~\bibnamefont
  {WU}},\ }\href {https://doi.org/10.1142/S0217984906012213} {\bibfield
  {journal} {\bibinfo  {journal} {Modern Physics Letters B}\ }\textbf {\bibinfo
  {volume} {20}},\ \bibinfo {pages} {1707} (\bibinfo {year} {2006})},\ \Eprint
  {https://arxiv.org/abs/https://doi.org/10.1142/S0217984906012213}
  {https://doi.org/10.1142/S0217984906012213} \BibitemShut {NoStop}%
\bibitem [{\citenamefont {Chen}\ \emph {et~al.}(2015)\citenamefont {Chen},
  \citenamefont {Xue}, \citenamefont {McCulloch}, \citenamefont {Chung},
  \citenamefont {Huang},\ and\ \citenamefont {Yip}}]{Chen2015}%
  \BibitemOpen
  \bibfield  {author} {\bibinfo {author} {\bibfnamefont {P.}~\bibnamefont
  {Chen}}, \bibinfo {author} {\bibfnamefont {Z.-L.}\ \bibnamefont {Xue}},
  \bibinfo {author} {\bibfnamefont {I.~P.}\ \bibnamefont {McCulloch}}, \bibinfo
  {author} {\bibfnamefont {M.-C.}\ \bibnamefont {Chung}}, \bibinfo {author}
  {\bibfnamefont {C.-C.}\ \bibnamefont {Huang}},\ and\ \bibinfo {author}
  {\bibfnamefont {S.-K.}\ \bibnamefont {Yip}},\ }\href
  {https://doi.org/10.1103/PhysRevLett.114.145301} {\bibfield  {journal}
  {\bibinfo  {journal} {Phys. Rev. Lett.}\ }\textbf {\bibinfo {volume} {114}},\
  \bibinfo {pages} {145301} (\bibinfo {year} {2015})}\BibitemShut {NoStop}%
\bibitem [{\citenamefont {Yang}\ and\ \citenamefont
  {Katsura}(2019{\natexlab{b}})}]{Yang2019}%
  \BibitemOpen
  \bibfield  {author} {\bibinfo {author} {\bibfnamefont {H.}~\bibnamefont
  {Yang}}\ and\ \bibinfo {author} {\bibfnamefont {H.}~\bibnamefont {Katsura}},\
  }\href {https://doi.org/10.1103/PhysRevLett.122.053401} {\bibfield  {journal}
  {\bibinfo  {journal} {Phys. Rev. Lett.}\ }\textbf {\bibinfo {volume} {122}},\
  \bibinfo {pages} {053401} (\bibinfo {year} {2019}{\natexlab{b}})}\BibitemShut
  {NoStop}%
\bibitem [{\citenamefont {Gorshkov}\ \emph {et~al.}(2010)\citenamefont
  {Gorshkov}, \citenamefont {Hermele}, \citenamefont {Gurarie}, \citenamefont
  {Xu}, \citenamefont {Julienne}, \citenamefont {Ye}, \citenamefont {Zoller},
  \citenamefont {Demler}, \citenamefont {Lukin},\ and\ \citenamefont
  {Rey}}]{Gorshkov2010}%
  \BibitemOpen
  \bibfield  {author} {\bibinfo {author} {\bibfnamefont {A.~V.}\ \bibnamefont
  {Gorshkov}}, \bibinfo {author} {\bibfnamefont {M.}~\bibnamefont {Hermele}},
  \bibinfo {author} {\bibfnamefont {V.}~\bibnamefont {Gurarie}}, \bibinfo
  {author} {\bibfnamefont {C.}~\bibnamefont {Xu}}, \bibinfo {author}
  {\bibfnamefont {P.~S.}\ \bibnamefont {Julienne}}, \bibinfo {author}
  {\bibfnamefont {J.}~\bibnamefont {Ye}}, \bibinfo {author} {\bibfnamefont
  {P.}~\bibnamefont {Zoller}}, \bibinfo {author} {\bibfnamefont
  {E.}~\bibnamefont {Demler}}, \bibinfo {author} {\bibfnamefont {M.~D.}\
  \bibnamefont {Lukin}},\ and\ \bibinfo {author} {\bibfnamefont
  {A.}~\bibnamefont {Rey}},\ }\href
  {https://doi.org/https://doi.org/10.1038/nphys1535} {\bibfield  {journal}
  {\bibinfo  {journal} {Nature physics}\ }\textbf {\bibinfo {volume} {6}},\
  \bibinfo {pages} {289} (\bibinfo {year} {2010})}\BibitemShut {NoStop}%
\bibitem [{\citenamefont {Cazalilla}\ \emph {et~al.}(2009)\citenamefont
  {Cazalilla}, \citenamefont {Ho},\ and\ \citenamefont {Ueda}}]{Cazalilla2009}%
  \BibitemOpen
  \bibfield  {author} {\bibinfo {author} {\bibfnamefont {M.~A.}\ \bibnamefont
  {Cazalilla}}, \bibinfo {author} {\bibfnamefont {A.~F.}\ \bibnamefont {Ho}},\
  and\ \bibinfo {author} {\bibfnamefont {M.}~\bibnamefont {Ueda}},\ }\href
  {https://doi.org/10.1088/1367-2630/11/10/103033} {\bibfield  {journal}
  {\bibinfo  {journal} {New Journal of Physics}\ }\textbf {\bibinfo {volume}
  {11}},\ \bibinfo {pages} {103033} (\bibinfo {year} {2009})}\BibitemShut
  {NoStop}%
\bibitem [{\citenamefont {Messio}\ and\ \citenamefont
  {Mila}(2012)}]{Messio2012}%
  \BibitemOpen
  \bibfield  {author} {\bibinfo {author} {\bibfnamefont {L.}~\bibnamefont
  {Messio}}\ and\ \bibinfo {author} {\bibfnamefont {F.}~\bibnamefont {Mila}},\
  }\href {https://doi.org/10.1103/PhysRevLett.109.205306} {\bibfield  {journal}
  {\bibinfo  {journal} {Phys. Rev. Lett.}\ }\textbf {\bibinfo {volume} {109}},\
  \bibinfo {pages} {205306} (\bibinfo {year} {2012})}\BibitemShut {NoStop}%
\bibitem [{\citenamefont {Cazalilla}\ and\ \citenamefont
  {Rey}(2014)}]{Cazalilla2014}%
  \BibitemOpen
  \bibfield  {author} {\bibinfo {author} {\bibfnamefont {M.~A.}\ \bibnamefont
  {Cazalilla}}\ and\ \bibinfo {author} {\bibfnamefont {A.~M.}\ \bibnamefont
  {Rey}},\ }\href {https://doi.org/10.1088/0034-4885/77/12/124401} {\bibfield
  {journal} {\bibinfo  {journal} {Reports on Progress in Physics}\ }\textbf
  {\bibinfo {volume} {77}},\ \bibinfo {pages} {124401} (\bibinfo {year}
  {2014})}\BibitemShut {NoStop}%
\bibitem [{\citenamefont {Zhang}\ \emph {et~al.}(2014)\citenamefont {Zhang},
  \citenamefont {Bishof}, \citenamefont {Bromley}, \citenamefont {Kraus},
  \citenamefont {Safronova}, \citenamefont {Zoller}, \citenamefont {Rey},\ and\
  \citenamefont {Ye}}]{Zhang2014}%
  \BibitemOpen
  \bibfield  {author} {\bibinfo {author} {\bibfnamefont {X.}~\bibnamefont
  {Zhang}}, \bibinfo {author} {\bibfnamefont {M.}~\bibnamefont {Bishof}},
  \bibinfo {author} {\bibfnamefont {S.~L.}\ \bibnamefont {Bromley}}, \bibinfo
  {author} {\bibfnamefont {C.~V.}\ \bibnamefont {Kraus}}, \bibinfo {author}
  {\bibfnamefont {M.~S.}\ \bibnamefont {Safronova}}, \bibinfo {author}
  {\bibfnamefont {P.}~\bibnamefont {Zoller}}, \bibinfo {author} {\bibfnamefont
  {A.~M.}\ \bibnamefont {Rey}},\ and\ \bibinfo {author} {\bibfnamefont
  {J.}~\bibnamefont {Ye}},\ }\href {https://doi.org/10.1126/science.1254978}
  {\bibfield  {journal} {\bibinfo  {journal} {Science}\ }\textbf {\bibinfo
  {volume} {345}},\ \bibinfo {pages} {1467} (\bibinfo {year}
  {2014})}\BibitemShut {NoStop}%
\bibitem [{\citenamefont {Taie}\ \emph {et~al.}(2022)\citenamefont {Taie},
  \citenamefont {Ibarra-Garc{\'\i}a-Padilla}, \citenamefont {Nishizawa},
  \citenamefont {Takasu}, \citenamefont {Kuno}, \citenamefont {Wei},
  \citenamefont {Scalettar}, \citenamefont {Hazzard},\ and\ \citenamefont
  {Takahashi}}]{Taie2022}%
  \BibitemOpen
  \bibfield  {author} {\bibinfo {author} {\bibfnamefont {S.}~\bibnamefont
  {Taie}}, \bibinfo {author} {\bibfnamefont {E.}~\bibnamefont
  {Ibarra-Garc{\'\i}a-Padilla}}, \bibinfo {author} {\bibfnamefont
  {N.}~\bibnamefont {Nishizawa}}, \bibinfo {author} {\bibfnamefont
  {Y.}~\bibnamefont {Takasu}}, \bibinfo {author} {\bibfnamefont
  {Y.}~\bibnamefont {Kuno}}, \bibinfo {author} {\bibfnamefont {H.-T.}\
  \bibnamefont {Wei}}, \bibinfo {author} {\bibfnamefont {R.~T.}\ \bibnamefont
  {Scalettar}}, \bibinfo {author} {\bibfnamefont {K.~R.}\ \bibnamefont
  {Hazzard}},\ and\ \bibinfo {author} {\bibfnamefont {Y.}~\bibnamefont
  {Takahashi}},\ }\href
  {https://doi.org/https://doi.org/10.1038/s41567-022-01725-6} {\bibfield
  {journal} {\bibinfo  {journal} {Nature Physics}\ }\textbf {\bibinfo {volume}
  {18}},\ \bibinfo {pages} {1356} (\bibinfo {year} {2022})}\BibitemShut
  {NoStop}%
\bibitem [{\citenamefont {Hilker}\ \emph {et~al.}(2017)\citenamefont {Hilker},
  \citenamefont {Salomon}, \citenamefont {Grusdt}, \citenamefont {Omran},
  \citenamefont {Boll}, \citenamefont {Demler}, \citenamefont {Bloch},\ and\
  \citenamefont {Gross}}]{Hilker2017}%
  \BibitemOpen
  \bibfield  {author} {\bibinfo {author} {\bibfnamefont {T.~A.}\ \bibnamefont
  {Hilker}}, \bibinfo {author} {\bibfnamefont {G.}~\bibnamefont {Salomon}},
  \bibinfo {author} {\bibfnamefont {F.}~\bibnamefont {Grusdt}}, \bibinfo
  {author} {\bibfnamefont {A.}~\bibnamefont {Omran}}, \bibinfo {author}
  {\bibfnamefont {M.}~\bibnamefont {Boll}}, \bibinfo {author} {\bibfnamefont
  {E.}~\bibnamefont {Demler}}, \bibinfo {author} {\bibfnamefont
  {I.}~\bibnamefont {Bloch}},\ and\ \bibinfo {author} {\bibfnamefont
  {C.}~\bibnamefont {Gross}},\ }\href {https://doi.org/10.1126/science.aam8990}
  {\bibfield  {journal} {\bibinfo  {journal} {Science}\ }\textbf {\bibinfo
  {volume} {357}},\ \bibinfo {pages} {484} (\bibinfo {year}
  {2017})}\BibitemShut {NoStop}%
\bibitem [{\citenamefont {Vijayan}\ \emph {et~al.}(2020)\citenamefont
  {Vijayan}, \citenamefont {Sompet}, \citenamefont {Salomon}, \citenamefont
  {Koepsell}, \citenamefont {Hirthe}, \citenamefont {Bohrdt}, \citenamefont
  {Grusdt}, \citenamefont {Bloch},\ and\ \citenamefont {Gross}}]{Vijayan2020}%
  \BibitemOpen
  \bibfield  {author} {\bibinfo {author} {\bibfnamefont {J.}~\bibnamefont
  {Vijayan}}, \bibinfo {author} {\bibfnamefont {P.}~\bibnamefont {Sompet}},
  \bibinfo {author} {\bibfnamefont {G.}~\bibnamefont {Salomon}}, \bibinfo
  {author} {\bibfnamefont {J.}~\bibnamefont {Koepsell}}, \bibinfo {author}
  {\bibfnamefont {S.}~\bibnamefont {Hirthe}}, \bibinfo {author} {\bibfnamefont
  {A.}~\bibnamefont {Bohrdt}}, \bibinfo {author} {\bibfnamefont
  {F.}~\bibnamefont {Grusdt}}, \bibinfo {author} {\bibfnamefont
  {I.}~\bibnamefont {Bloch}},\ and\ \bibinfo {author} {\bibfnamefont
  {C.}~\bibnamefont {Gross}},\ }\href {https://doi.org/10.1126/science.aay2354}
  {\bibfield  {journal} {\bibinfo  {journal} {Science}\ }\textbf {\bibinfo
  {volume} {367}},\ \bibinfo {pages} {186} (\bibinfo {year}
  {2020})}\BibitemShut {NoStop}%
\bibitem [{\citenamefont {Emery}\ \emph {et~al.}(1990)\citenamefont {Emery},
  \citenamefont {Kivelson},\ and\ \citenamefont {Lin}}]{Emery1990}%
  \BibitemOpen
  \bibfield  {author} {\bibinfo {author} {\bibfnamefont {V.~J.}\ \bibnamefont
  {Emery}}, \bibinfo {author} {\bibfnamefont {S.~A.}\ \bibnamefont
  {Kivelson}},\ and\ \bibinfo {author} {\bibfnamefont {H.~Q.}\ \bibnamefont
  {Lin}},\ }\href {https://doi.org/10.1103/PhysRevLett.64.475} {\bibfield
  {journal} {\bibinfo  {journal} {Phys. Rev. Lett.}\ }\textbf {\bibinfo
  {volume} {64}},\ \bibinfo {pages} {475} (\bibinfo {year} {1990})}\BibitemShut
  {NoStop}%
\bibitem [{\citenamefont {Ogata}\ \emph {et~al.}(1991)\citenamefont {Ogata},
  \citenamefont {Luchini}, \citenamefont {Sorella},\ and\ \citenamefont
  {Assaad}}]{Ogata1991}%
  \BibitemOpen
  \bibfield  {author} {\bibinfo {author} {\bibfnamefont {M.}~\bibnamefont
  {Ogata}}, \bibinfo {author} {\bibfnamefont {M.~U.}\ \bibnamefont {Luchini}},
  \bibinfo {author} {\bibfnamefont {S.}~\bibnamefont {Sorella}},\ and\ \bibinfo
  {author} {\bibfnamefont {F.~F.}\ \bibnamefont {Assaad}},\ }\href
  {https://doi.org/10.1103/PhysRevLett.66.2388} {\bibfield  {journal} {\bibinfo
   {journal} {Phys. Rev. Lett.}\ }\textbf {\bibinfo {volume} {66}},\ \bibinfo
  {pages} {2388} (\bibinfo {year} {1991})}\BibitemShut {NoStop}%
\bibitem [{\citenamefont {Grusdt}\ and\ \citenamefont
  {Pollet}(2020)}]{Grusdt2020}%
  \BibitemOpen
  \bibfield  {author} {\bibinfo {author} {\bibfnamefont {F.}~\bibnamefont
  {Grusdt}}\ and\ \bibinfo {author} {\bibfnamefont {L.}~\bibnamefont
  {Pollet}},\ }\href {https://doi.org/10.1103/PhysRevLett.125.256401}
  {\bibfield  {journal} {\bibinfo  {journal} {Phys. Rev. Lett.}\ }\textbf
  {\bibinfo {volume} {125}},\ \bibinfo {pages} {256401} (\bibinfo {year}
  {2020})}\BibitemShut {NoStop}%
\bibitem [{\citenamefont {Hirthe}\ \emph {et~al.}(2023)\citenamefont {Hirthe},
  \citenamefont {Chalopin}, \citenamefont {Bourgund}, \citenamefont
  {Bojovi{\'c}}, \citenamefont {Bohrdt}, \citenamefont {Demler}, \citenamefont
  {Grusdt}, \citenamefont {Bloch},\ and\ \citenamefont {Hilker}}]{Hirthe2023}%
  \BibitemOpen
  \bibfield  {author} {\bibinfo {author} {\bibfnamefont {S.}~\bibnamefont
  {Hirthe}}, \bibinfo {author} {\bibfnamefont {T.}~\bibnamefont {Chalopin}},
  \bibinfo {author} {\bibfnamefont {D.}~\bibnamefont {Bourgund}}, \bibinfo
  {author} {\bibfnamefont {P.}~\bibnamefont {Bojovi{\'c}}}, \bibinfo {author}
  {\bibfnamefont {A.}~\bibnamefont {Bohrdt}}, \bibinfo {author} {\bibfnamefont
  {E.}~\bibnamefont {Demler}}, \bibinfo {author} {\bibfnamefont
  {F.}~\bibnamefont {Grusdt}}, \bibinfo {author} {\bibfnamefont
  {I.}~\bibnamefont {Bloch}},\ and\ \bibinfo {author} {\bibfnamefont {T.~A.}\
  \bibnamefont {Hilker}},\ }\href
  {https://doi.org/https://doi.org/10.1038/s41586-022-05437-y} {\bibfield
  {journal} {\bibinfo  {journal} {Nature}\ }\textbf {\bibinfo {volume} {613}},\
  \bibinfo {pages} {463} (\bibinfo {year} {2023})}\BibitemShut {NoStop}%
\end{thebibliography}%


%


\end{document}